\documentclass[fleqn,usenatbib]{mnras}

\usepackage{newtxtext,newtxmath}

\usepackage[T1]{fontenc}
\usepackage{ae,aecompl}

\usepackage{graphicx}	
\usepackage{amsmath}	
\usepackage{amssymb}	

\title[$\alpha$-rich young stars]{Origin of $\alpha$-rich young  stars: clues from C, N and O}

\author[S. Hekker et al.]{
Saskia Hekker$^{1,2}$\thanks{E-mail: hekker@mps.mpg.de} and
Jennifer A. Johnson$^{3}$
\\
$^{1}$Max Planck Institute for Solar System Research, Justus-von-Liebig-Weg 3, 37077 G\"{o}ttingen, Germany. \\
$^{2}$Stellar Astrophysics Centre, Department of Physics and Astronomy, Aarhus University, Ny Munkegade 120, DK-8000 Aarhus C, Denmark \\  
$^{3}$Department of Astronomy and Center for Cosmology and Astroparticle Physics, Ohio State University, Columbus, OH 43210
}

\date{Accepted XXX. Received YYY; in original form ZZZ}

\pubyear{2019}

\begin{document}
\label{firstpage}
\pagerange{\pageref{firstpage}--\pageref{lastpage}}
\maketitle

\begin{abstract}
A small set of chemically old stars that appear young by their independently derived masses has been detected. These are so-called $\alpha$-rich young stars. For a sample of 51 red-giant stars, for which spectra are available from SDSS/APOGEE and masses are available from asteroseismic measures based on \textit{Kepler} lightcurves, we derive the C, N and O abundances through an independent analysis. These stars span a wide range of N/C surface number density ratios. We interpret the high-mass stars with low N/C as being products of mergers or mass transfer during or after first dredge up, because the dredge-up features are the same as for low-mass stars. The $\alpha$-rich young stars with high N/C follow the expected trend of N/C for their mass, and could be either genuine young stars (leaving their high [$\alpha$/Fe] unexplained) or the results of mergers on the main sequence.
\end{abstract}




\section{Introduction} 

Shortly after the discovery of metal-poor stars \citep{chamberlain1951}, these stars were found to have higher Mg/Fe, Si/Fe, and Ca/Fe than the Sun \citep{aller1960,wallerstein1962}. Mg, Si, and Ca are `$\alpha$ elements', a sequence of elements formed by the addition or subtraction of $\alpha$ particles. These $\alpha$ elements are ejected primarily in the core-collapse supernovae (ccSNe) of massive stars.
In contrast, Fe has in addition to the contribution from ccSNe substantial contributions from Type Ia SNe, whose progenitors are considerably longer-lived. \citet{tinsley1979} showed that therefore stars enhanced in [$\alpha/$Fe]\footnote{We use the standard notation: [X/Q]$=\,\mathrm{log}N{(X)}_{* }-\mathrm{log}N{(X)}_{\odot}\,-\mathrm{log}N{(Q)}_{* }+\mathrm{log}N{(Q)}_{\odot }$. X/Q refers to N(X)/N(Q).} are formed early in the chemical evolution of a system, when 
the production of iron lagged behind that of the $\alpha$ elements. 

Many studies have confirmed that $\alpha$-rich stars are in general old \citep[e.g.,][]{edvardsson1993,fuhrmann1998,vandenberg2002}. Two recent papers have provided an updated view of the $\alpha$-age relationship based on improved luminosities for stars that have been measured using parallaxes from the Gaia mission. \citet{Feuillet2018} combined luminosities with temperatures and gravities from spectroscopy to derive masses, and therefore ages, for first ascent red giant branch stars. \citet{delgado2019} compared the positions on the Herzsprung-Russell (HR) diagram of turnoff and subgiant stars with theoretical isochrones to derive ages. In both cases, they found that essentially all stars with [$\alpha$/Fe]$> 0.2 $ and [Fe/H]$ < -0.5$ are older than 10 Gyr.  

With the advent of asteroseismology, it became possible to determine stellar ages for large numbers of field subgiant and red-giant stars. Subgiants and red giants have run out of hydrogen in their cores. Because of the hydrogen exhaustion their evolution timescale is a sensitive function of mass and composition, and a measurement of both yields ages. The seismic signals are primarily related to the stellar mass and radius, while compositions are derived from high-resolution spectra. Catalogues combining asteroseismic and spectroscopic information now provide ages for thousands of field giants \citep[e.g.,][]{pinsonneault2014,pinsonneault2018,anders2017}.

These asteroseismic measurements led to the detection of $\alpha$-rich young stars \citep{chiappini2015, martig2015}, with [$\alpha$/Fe] > 0.13 dex and ages $<$ 6 Gyr as defined by \citet{martig2015}. \citet{chiappini2015} suggested that these stars may be formed near the co-rotation region of the Galactic bar where complex chemical evolution is expected. However, because the signature of youth is based on a relatively high mass on the red-giant branch, mass transfer from a binary companion or the merger of two stars would make a star more massive and hence appear young in asteroseismic analyses \citep{chiappini2015,martig2015}. There is a substantial population of $\alpha$-rich stars in the Milky Way that could provide the needed binary systems. This population is also known as the chemically defined thick disk, which is also characterised by older ages \citep[e.g.,][]{silvaaguirre2018}, and slower rotational velocity and higher vertical extent than the thin disk \citep[e.g.,][]{bensby2003}. 

Abundance analyses by \citet{yong2016, matsuno2018} show that the majority of $\alpha$-rich young stars have abundances in line with thick disk stars. Radial velocity measurements by \citet{jofre2016, matsuno2018} show that more than 50\% of their observed $\alpha$-rich young stars show substantial radial velocity scatter, interpreted as due to multiplicity, compared to a much lower fraction in comparison samples. Furthermore, \citet{tayar2015} found that one of these stars was rapidly rotating, another sign that an interaction could be involved. Together with the theoretical work by \citet{izzard2018} there is mounting evidence that at least a significant part of the massive, i.e. young $\alpha$-rich stars is a result of binary interactions. These interactions can either produce a higher mass star in a binary through mass transfer, or a more massive single star through a merger. However, one needs to note that the main conclusions of \citet{izzard2018} are a prediction that the fraction of thick disk giants with mass in excess of 1.3~M$_{\odot}$ ranges from 0.8 to 3 percent in all but one of their model sets. In contrast, \citet{martig2015} found $\sim$6\% of their $\alpha$-rich stars were high mass. Only when \citet{izzard2018} change from a log-normal distribution of orbital separations to a (less representative) logarithmically-flat separation distribution do they predict a fraction of giants with mass in excess of 1.3~M$_{\odot}$ to be large at 11\%. 

\citet{izzard2018} also included an analysis of [C/N] in their analysis of theoretical models. In these models they boosted their initial [C/N] by +0.2 dex to mimic chemical evolution in the thick disk. Additionally, they assume that merged stars have first dredge-up equally deep as in a single star of the same mass. In general \citet{izzard2018} find that interaction in binary-stars systems explain the observed range in surface [C/N], with stars with [C/N]~<~$-0.4$ almost all being merged binaries. 

In this work we revisit the trends in CNO abundances measured in the stellar atmospheres. The reason for studying these elements is that these are involved in the fusion reactions in the core and subsequently dredged up when the convection zone deepens in the early phases of the red-giant branch. If there are stars that are results of mergers, we may expect there to be some impact on the core burning and/or the depth of the dredge-up, which these surface abundances give a view upon. We first discuss the data used and the spectroscopic analysis that we performed. We then show the results and discuss possible implications.

\section{Data}
In this paper we present a reanalysis of the 26 stars from \cite{jofre2016}. These 26 stars comprise 13 young stars with APOGEE DR12 \citep{holtzman2015} [$\alpha$/M]~>~0.13~dex of \citet{martig2015}, which have masses above 1.4~M$_{\odot}$, and for each of them a `twin' star with similar atmospheric parameters and masses below 1.2~M$_{\odot}$. In addition to this set of stars, we extended the sample with 14 $\alpha$-rich young stars and 11 old stars selected from the APOKASC sample \citep{pinsonneault2018}. For the latter APOKASC sample we selected $\alpha$-rich young stars to have [$\alpha$/Fe]~>~0.2 and an age younger than 5~Gyr based on the results presented by \citep{pinsonneault2018}. The old stars were chosen to have [$\alpha$/Fe]~>~0.2 and to be older than 7~Gyr. These stars cover ranges in {\it T}$_{\rm eff}$, log~${g}$ and [Fe/H] of 4300$-$5000 K, 1.6$-$3.3 dex and $+$0.05 to $-$0.60 dex, respectively.

We performed an independent analysis (see Section 3) of the complete set of stars based on APOGEE spectra \citep{holtzman2018}. These spectra have been obtained by the second generation of the Apache Point Observatory Galactic Evolution Experiment (APOGEE-2, \citealt{majewski2017}) survey, as part of the fourth phase of Sloan Digital Sky Survey (SDSS-IV, \citealt{blanton2017}). APOGEE-2 spectra were obtained using the 2.5-m Sloan Foundation Telescope \citep{gunn2006} of the Apache Point Observatory in New Mexico, USA and cover a narrow wavelength band in the near infrared ($\lambda$$\sim$1.51$-$1.70 $\mu$m in H-band) with a high resolution ($R$$\sim$22,500) \citep{wilson2019}.

\section{Spectroscopic analysis}
We performed a standard spectroscopic abundance analysis adopting ATLAS9 photospheres assuming local thermodynamical equilibrium (LTE), Kurucz grid of models \citep{caskur2003}, a common line list (see Table~\ref{linelist}) and the LTE line analysis and spectrum synthesis code {\sc \bf MOOG} \citep{sneden1973}.  We performed an independent analysis of C, N and O abundances for several reasons. We wanted to include more lines than the automated pipeline \citep[ASPCAP,][]{holtzman2018} and to make sure that the continuum was placed in the best possible way. That is the spectrum was normalised interactively to unity using {\sc IRAF} by marking the continuum points, which were then fitted by a lower order (typically 3 to 5) cubic spline polynomial. However, the main reason for the present homogeneous study is to ensure that our abundances are on an excellent relative abundance scale to compare abundances of $\alpha$-rich young stars with ones of $\alpha$-rich old stars as well as $\alpha$-normal stars.

\begin{table}
\centering
\caption{Line list of neutral atomic lines selected for deriving chemical abundances. The wavelength ($\lambda$), lower excitation potential ($\chi$), oscillator strength (expressed as log $gf$) and damping constant are taken from \citet{melendez1999}.}
\label{linelist}
\begin{tabular}{lccrc}   
\hline
Element & $\lambda$ [\AA\ ] & $\chi$ [eV] & log $gf$ & damping constant\\
\hline
Fe & 15176.72 & 5.92  &  $-0.95$ & 1.1e$-$30\\
Fe & 15301.56 & 5.92  &  $-0.84$  & 1.0e$-$30\\
Fe & 15394.67 & 5.62  &  $-0.28$ &  7.6e$-$31\\
Fe & 15395.72 & 5.62  &  $-0.41$ &  7.6e$-$31\\  
Fe & 15485.45 & 6.28  &  $-0.93$ &  3.1e$-$30\\
Fe & 15524.30 & 5.79  &  $-1.51$ &  6.5e$-$31\\     
Fe & 15534.26 & 5.64  &  $-0.47$ &  9.0e$-$31\\       
Fe & 15648.52 & 5.43  &  $-0.80$ &  6.3e$-$31\\  
Fe & 15652.87 & 6.25  &  $-0.19$ &  1.5e$-$30\\  
Fe & 15733.51 & 6.25  &  $-0.76$ &  1.5e$-$30\\
Fe & 15761.31 & 6.25  &  $-0.23$  &  1.5e$-$30\\
Fe & 15774.07 & 6.39  &     0.25   &  2.1e$-$30\\
Fe & 15609.04 & 5.62  &  $-0.34$ &  8.8e$-$31\\
Fe & 15904.35 & 6.36  &     0.25   &  3.0e$-$30\\ 
Fe & 15906.04 & 5.62  &  $-0.34$ &  8.8e$-$31\\
Fe & 15934.02 & 6.31  &  $-0.43$ &  2.1e$-$30\\   
Fe & 15952.63 & 6.34  &  $-0.81$ &  3.8e$-$30\\ 
Fe & 15964.87 & 5.92  &  $-0.23$ & 1.1e$-$30\\
Fe & 15980.73 & 6.26  &    0.60 &  1.5e$-$30\\
Fe & 15997.74 & 5.92  &  $-0.63$ &  1.0e$-$30\\ 
Fe & 16040.65 & 5.87  &  $-0.07$ &  9.9e$-$31\\  
Fe & 16102.41 & 5.87  &   $0.08$ &  9.9e$-$31\\    
Fe & 16180.93 & 6.28  &   0.08     &  1.5e$-$30\\  
Fe & 16198.51 & 5.41  &  $-0.60$ &  5.6e$-$31\\
Fe & 16225.64 & 6.38  &  $-0.03$ &  2.9e$-$30\\   
Fe & 16292.85 & 5.92  &  $-0.62$ & 1.0e$-$30\\  
Fe & 16324.46 & 5.39  &  $-0.66$ & 5.3e$-$31\\   
Fe & 16331.53 & 5.98  &  $-0.61$ & 1.1e$-$30\\
Ni & 16310.51 &  5.28  &  $-0.12$ & 7.1e$-$31\\ 
Ni & 16363.11 &  5.28  &   0.28     & 7.1e$-$31\\  
Ni & 16388.75 & 6.03  &   $-0.27$ & 3.1e$-$29\\
Ni & 16584.44 &  5.30  &  $-0.78$ & 7.4e$-$31\\ 
Ni & 16673.71 &  6.03  &    0.21    & 2.9e$-$30\\     
Ni & 16815.47 & 5.30  &   $-0.70$ & 7.2e$-$31\\
Mg & 15879.56 & 5.95  &  $-1.39$ & 2.7e$-$30\\ 
Mg & 15886.19 & 5.95  &  $-1.66$ &  2.7e$-$30\\
Al & 16718.98   & 4.09  &   0.19  & 3.0e$-$31\\   
Al & 16750.57 & 4.09 & 0.50 & 3.0e$-$31\\
Al & 16763.35   & 4.09  &  $-0.64$  & 2.0e$-$30\\   
Si & 15557.79    & 5.96  &  $-0.90$ &  2.0e$-$30\\
Si & 16060.02    & 5.95  &  $-0.66$ & 2.0e$-$30\\  
Si & 16163.71  & 5.95 &   $-0.99$ &  4.4e$-$31\\  
Si & 16215.71  & 5.95 &  $-0.66$  &   4.3e$-$31\\
Si & 16241.87  & 5.96 &  $-0.87$  & 4.4e$-$31\\
Si & 16828.18   & 5.98  &  $-1.26$ &  4.4e$-$31\\ 
Ca &16150.76   & 4.53  &  $-0.34$ &  3.1e$-$30\\ 
Ca &16155.27   & 4.53  &  $-0.80$ & 3.0e$-$30\\ 
Ca & 16197.04  & 4.53  & $-0.02$ & 3.0e$-$30\\   
Ti & 15602.84   & 2.27  &  $-1.81$ &  4.9e$-$32\\   
Ti & 15715.57  & 1.87  &  $-1.59$ & 3.6e$-$32\\    
Ti & 15543.78  &  1.88  &  $-1.48$ &  3.6e$-$32\\   
\hline
\end{tabular}
\end{table}

\subsection{Spectral line selection}
The crucial aspect in deriving accurate elemental abundances is the selection of suitable stellar lines with good quality atomic and molecular data\footnote{see for example, \url{http://kurucz.harvard.edu/linelists.html}}. An infrared solar spectral atlas, with a resolution of 0.05 \AA\ \citep{livingston1991} observed with the Fourier Transform Spectrometer at the McMath/Pierce Solar Telescope on Kitt Peak \citep{pierce1964} and  degraded to the instrumental resolution of the APOGEE spectrograph, was used as a reference to visually confirm clean, unblended, relatively isolated and symmetric spectral lines due to various atomic and molecular species. The suitable spectral lines of an element are those lines which contain only contributions from the element at the central wavelength specified by the transition. The equivalent width (EW) of the selected atomic absorption lines in red giants were measured manually using the cursor commands in \textit{splot} package of {\sc IRAF} by fitting often a Gaussian profile. For a few lines, i.e. the 15060~\AA\ and 16215~\AA\ lines of Si and the 16719~\AA\ and 16763~\AA\ lines of Al, which are typically strong with EW > 200~m\AA\ and have a Lorentzian shaped profile, a direct integration was performed for the best measure of EW. The abundances of carbon, nitrogen, oxygen were derived by matching the synthetic spectra to selected regions around CO (15575 - 15585~\AA\ ), CN (15370 - 15420~\AA\ ; 15462 - 15472~\AA\ ; 15560 - 15590~\AA\ ) and OH (16367 - 16369~\AA\ ) lines in the observed spectrum.

The oscillator strengths ($f$) for the elements Fe, Ni, Mg, Al, Si, Ca and Ti in the infrared H-band were extracted from \cite{melendez1999}. These oscillator strengths were derived by \citet{melendez1999} from an iterative fit of synthetic spectra to the solar spectrum. The typical accuracies in the adopted log\,$gf$-values (where $g$ is the statistical weight of an atomic energy level) are better than 0.05 dex.  The molecular data for OH were extracted from \cite{brooke2016}, CO and CN molecular data were taken from the Kurucz coxx.asc\footnote{http://kurucz.harvard.edu/linelists/linesmol/} line list and \cite{sneden2014}, respectively.

Although the spectra of red giants at the spectral resolution of the APOGEE spectrograph in the H-band are heavily crowded with absorption lines, our line selection results in a number of unblended lines for each element. The final list of neutral atomic lines selected for deriving chemical abundances includes Fe(28), Ni(6), Mg(2), Al(3), Si(6), Ca(3), Ti(3). Here, the numbers in the parentheses indicate the number of lines of an element considered in the abundance analysis. The full line list used in this work is shown in Table~\ref{linelist}, while the measured EWs for all stars are listed in Tables~\ref{EWs0} - \ref{EWs6}. Additionally, we used a set of molecular lines due to CN (at wavelength 15374, 15400, 15466, 15576, 15581 \AA\ ), OH at 16368 \AA\ and a CO molecular band head at 15578 \AA. The molecular line list employed here involves the dominant isotopes of each element and is composed of $^{12}$C$^{16}$O, $^{12}$C$^{14}$N and $^{16}$O$^{1}$H molecules.

\begin{figure*}
\centering
\begin{minipage}{0.32\linewidth}
\includegraphics[width=\linewidth]{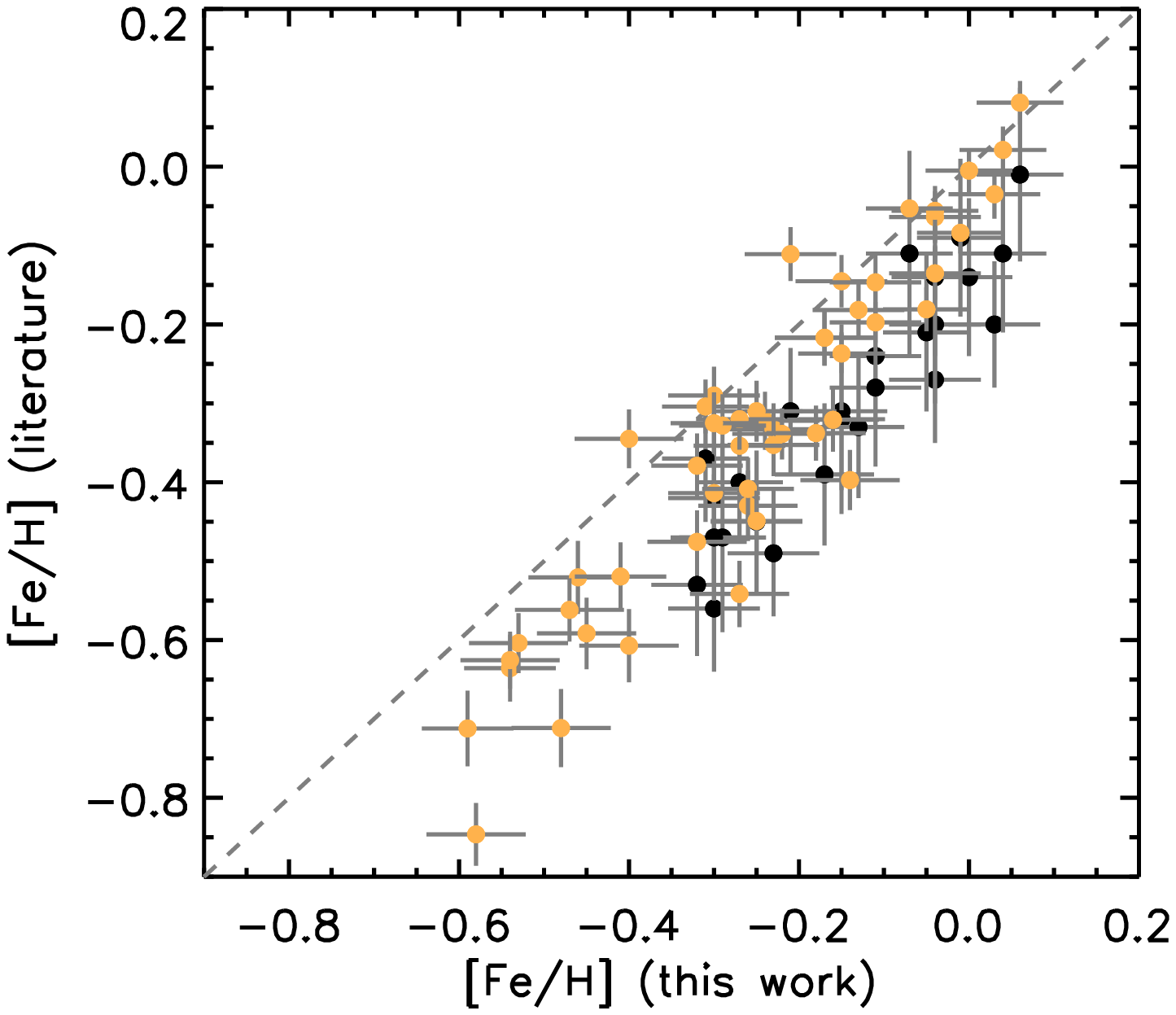}
\end{minipage}
\begin{minipage}{0.32\linewidth}
\includegraphics[width=\linewidth]{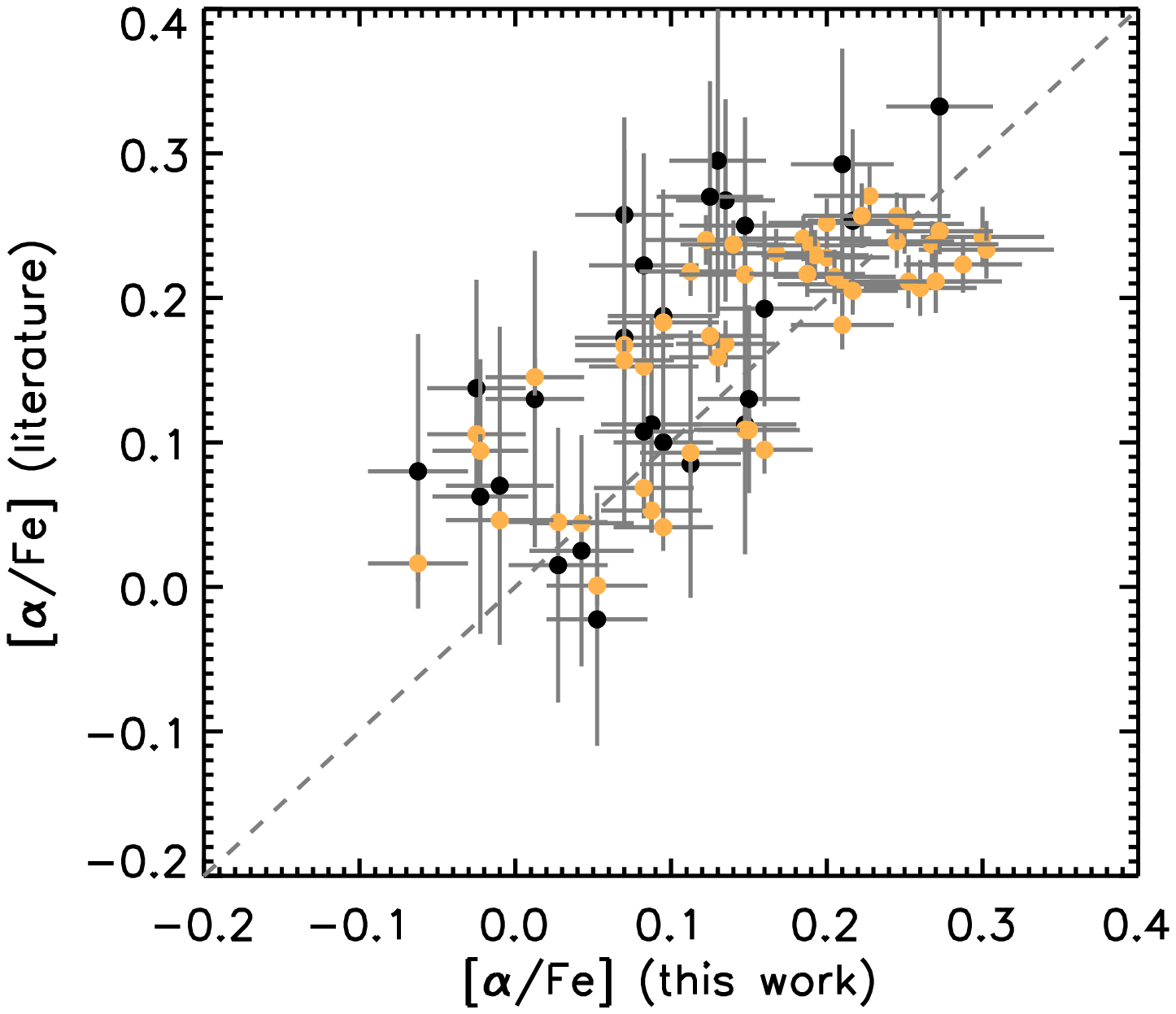}
\end{minipage}
\begin{minipage}{0.32\linewidth}
\includegraphics[width=\linewidth]{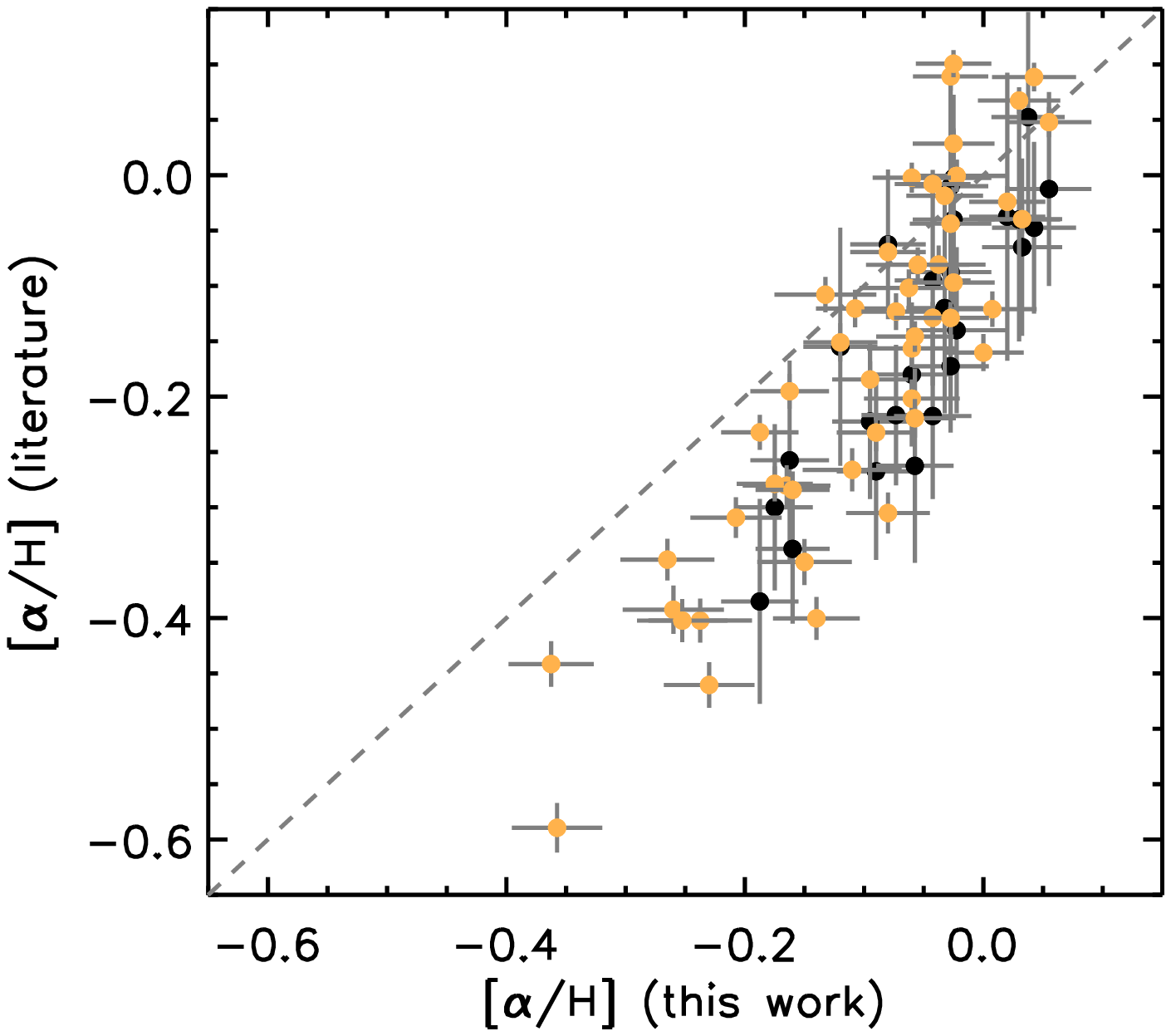}
\end{minipage}
\begin{minipage}{0.32\linewidth}
\includegraphics[width=\linewidth]{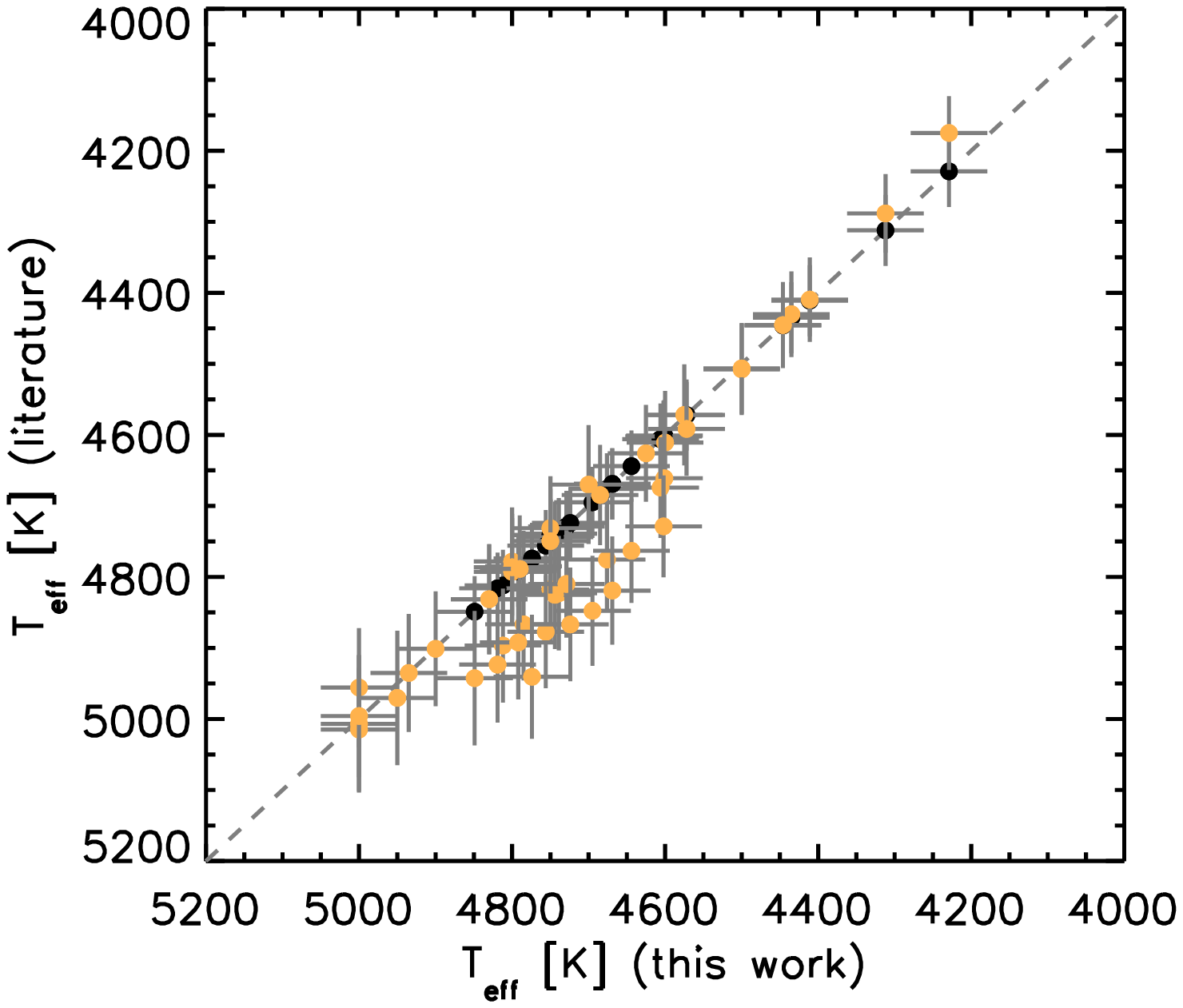}
\end{minipage}
\begin{minipage}{0.32\linewidth}
\includegraphics[width=\linewidth]{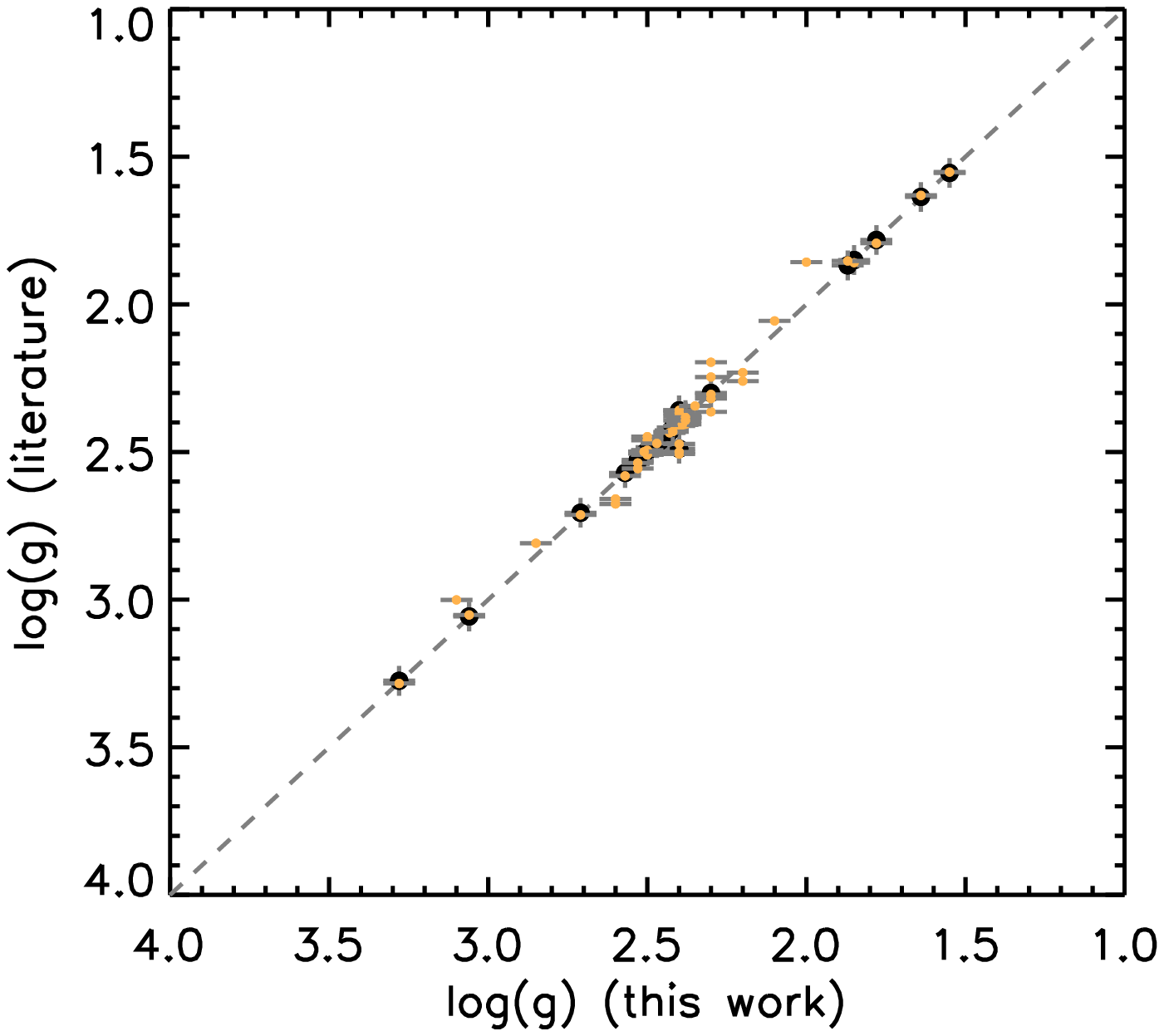}
\end{minipage}
\begin{minipage}{0.32\linewidth}
\includegraphics[width=\linewidth]{empty}
\end{minipage}
\caption{Comparisons between [Fe/H] results (left, top), [$\alpha$/Fe] results (centre, top),  [$\alpha$/H] (right), $T_{\rm eff}$ (left, bottom) and log~$g$ (centre, bottom) obtained in this work and in the literature. Orange dots show comparisons with results presented in the APOGEE DR14 \citep{holtzman2018} and black points show comparisons with results obtained with the BACCHUS pipeline \citep{hawkins2016}. The gray dashed line indicates the one-to-one relation. The uncertainties for [$\alpha$/Fe] and [$\alpha$/H] for the BACCHUS sample and this work were determined by calculating the [$\alpha$/X] with all individual [Mg, Si, Ca, Ti/X] values set to their 1-$\sigma$ upper or lower limits. This conservatively assumes maximum correlation among the measurements of the individual elements. For APOGEE DR14 results, we use the quoted uncertainties from DR14, which reflect the scatter in the [$\alpha$/Fe] ratios in clusters of stars \citep{holtzman2018}. For $T_{\rm eff}$ and log~$g$ uncertainties for the BACCHUS pipeline and our work are estimated to be 50~K and 0.05~dex respectively, while the uncertainties for APOGEE DR14 are taken from that database.}
\label{comp_fealpha}
\end{figure*}

\subsection{Solar abundances}
The solar abundances were derived using solar EWs, measured off the infrared solar spectral atlas \citep{livingston1991}, and the ATLAS9 theoretical photosphere computed adopting the solar abundances from \cite{asplund2006} with parameters {\it T}$_{\rm eff \odot}$ = 5777 K, log~${g}_{\odot}$ = 4.44 dex, [Fe/H]=0.0 dex to establish a reference abundance scale. Lines with individual iron abundances falling outside of $\pm0.05$ dex of the mean log\,$\epsilon$(Fe)=7.47 dex were rejected assuming either the selected Fe lines have inadequate atomic data or are blended with unidentified lines. This procedure is repeated for lines of other atomic species. We obtained a microturbulence velocity of $\xi_{t}$ = 0.97 km s$^{-1}$, a value in agreement with the published results \citep{asplund2006}, by requiring that the Fe abundance from \ion{Fe}{1} lines is independent of the EW of the line.  

The derived solar abundances are reported in Tables \ref{chemical_abundances} and \ref{cno_abundances}. We refer to our solar abundances when determining the stellar abundances, [X/H] and [X/Fe], i.e., our analysis is essentially a differential one relative to the Sun.   

\subsection{Stellar abundances}
The EW of a spectral line is influenced by the physical conditions and number density of absorbers in the stellar atmosphere. Therefore, it is essential to predetermine the stellar parameters to obtain a robust estimate of chemical abundances. 
Starting with a theoretical model with {\it T}$_{\rm eff}$ derived from the ASPCAP pipeline \citep{holtzman2018}, asteroseismic log~${g}$ \citep[][for the stars analysed by \citet{jofre2016} and the newly selected stars in the current work, respectively]{pinsonneault2014,pinsonneault2018} and measured EWs, the individual line abundances were obtained by matching the computed EWs to the observed ones \citep{sneden1973} while satisfying the following constraints simultaneously. First, we obtain the microturbulence, $\xi_{t}$, assumed to be isotropic and depth independent, by requiring that the Fe abundance from \ion{Fe}{1} lines is independent of the EW of the line. Second, we estimate the effective temperature by requiring that the abundance from \ion{Fe}{1} lines is independent of the lower excitation potential of the line (excitation equilibrium). As the stellar parameters {\it T}$_{\rm eff}$, log~${g}$ and $\xi_{\rm t}$ are interdependent, several iterations are needed to choose a suitable model from the grid. If required for convergence the asteroseismic log~${g}$ values were also iterated over within their uncertainties to satisfy that the \ion{Fe}{I} lines are independent of both the EW and the lower excitation potential simultaneously. No \ion{Fe}{II} lines are present in the APOGEE wavelength regime and hence ionisation balance could not be enforced. The stellar parameters ($T_{\rm eff}$ and log $g$) used in the current analysis are listed in Table~\ref{stellarproperties} together with the masses, radii and ages as presented by \citet{pinsonneault2018}.

\begin{figure*}
\centering
\begin{minipage}{0.24\linewidth}
\includegraphics[width=\linewidth]{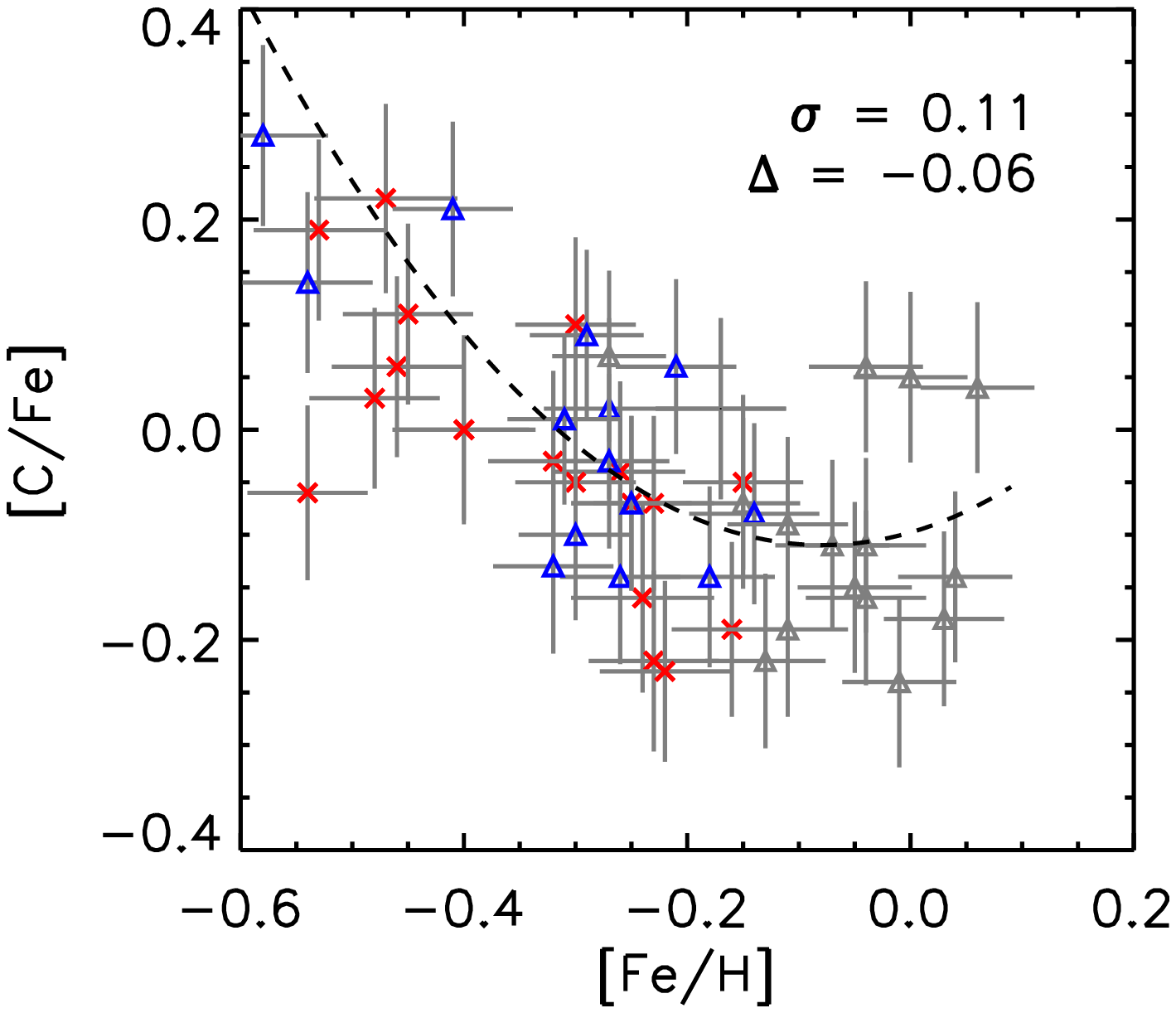}
\end{minipage}
\begin{minipage}{0.24\linewidth}
\includegraphics[width=\linewidth]{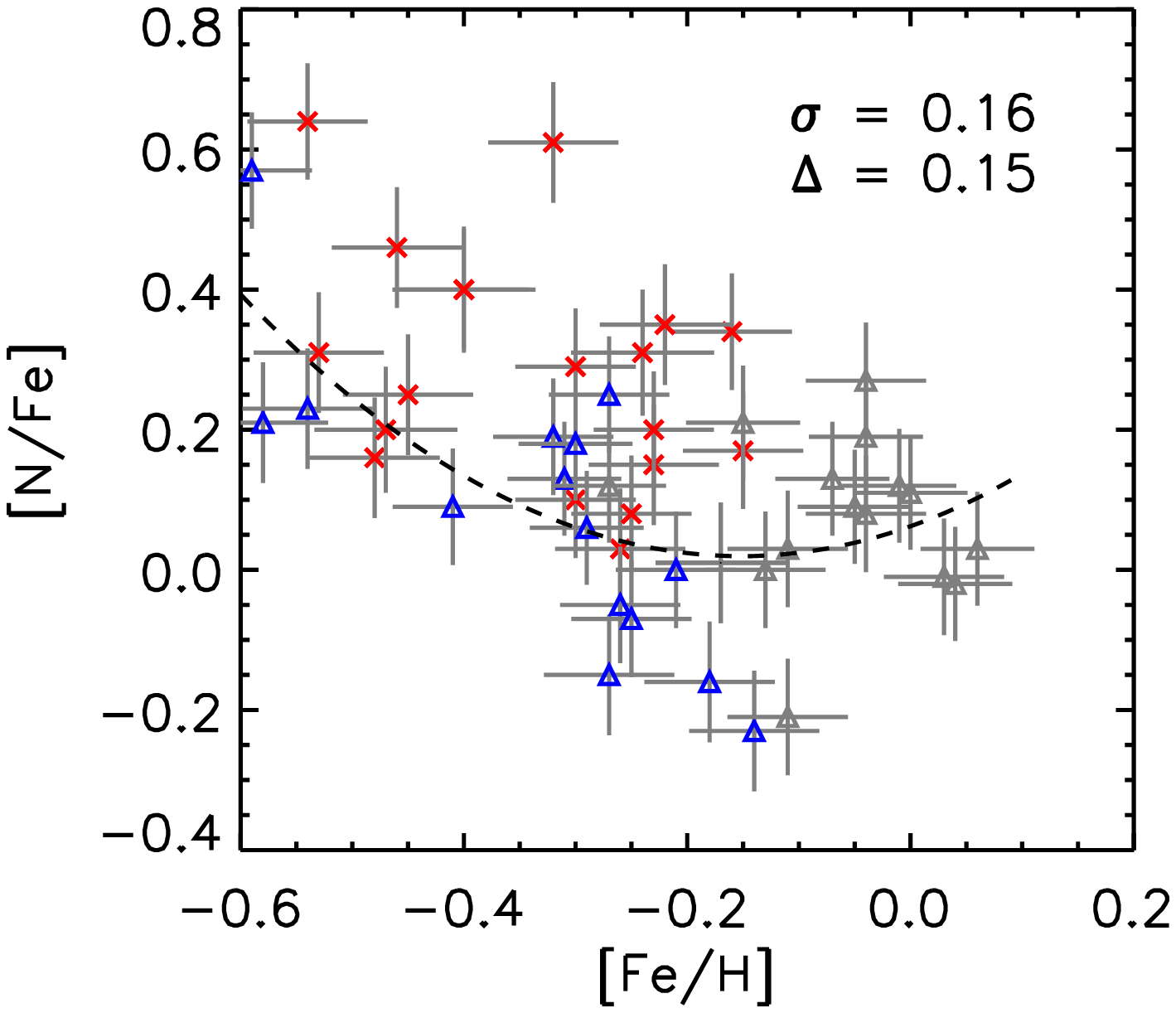}
\end{minipage}
\begin{minipage}{0.24\linewidth}
\includegraphics[width=\linewidth]{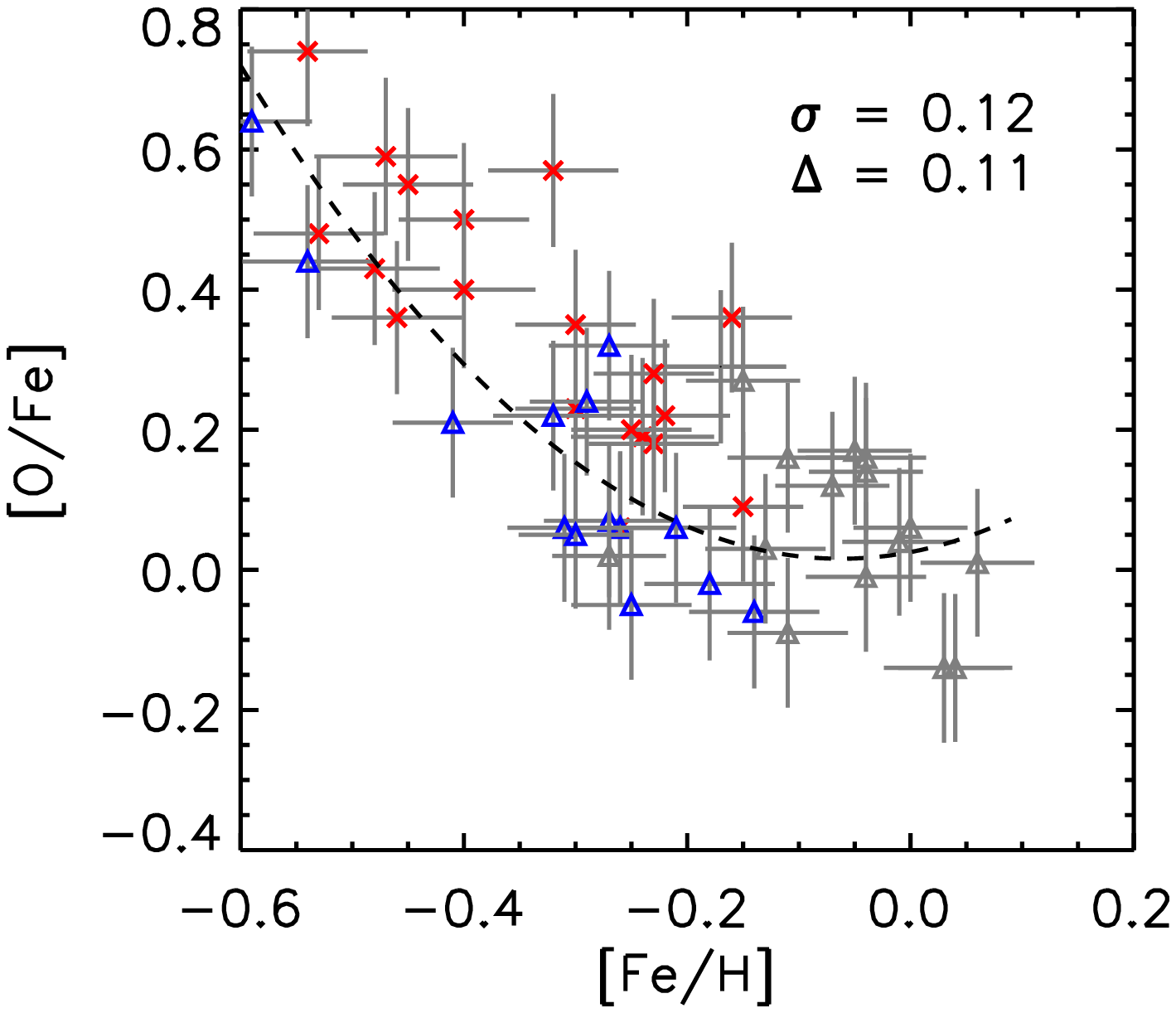}
\end{minipage}
\begin{minipage}{0.24\linewidth}
\includegraphics[width=\linewidth]{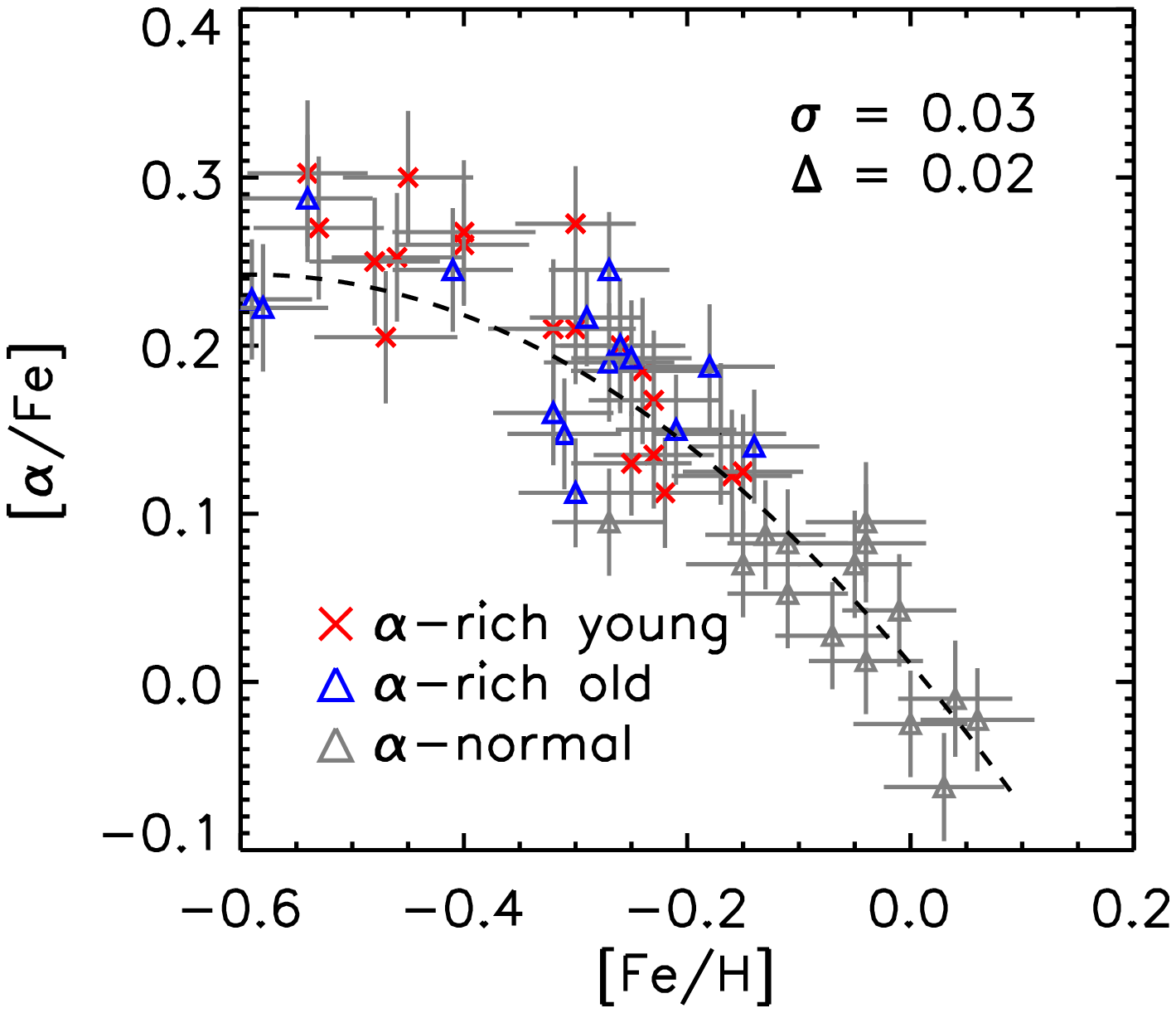}
\end{minipage}
\begin{minipage}{0.24\linewidth}
\includegraphics[width=\linewidth]{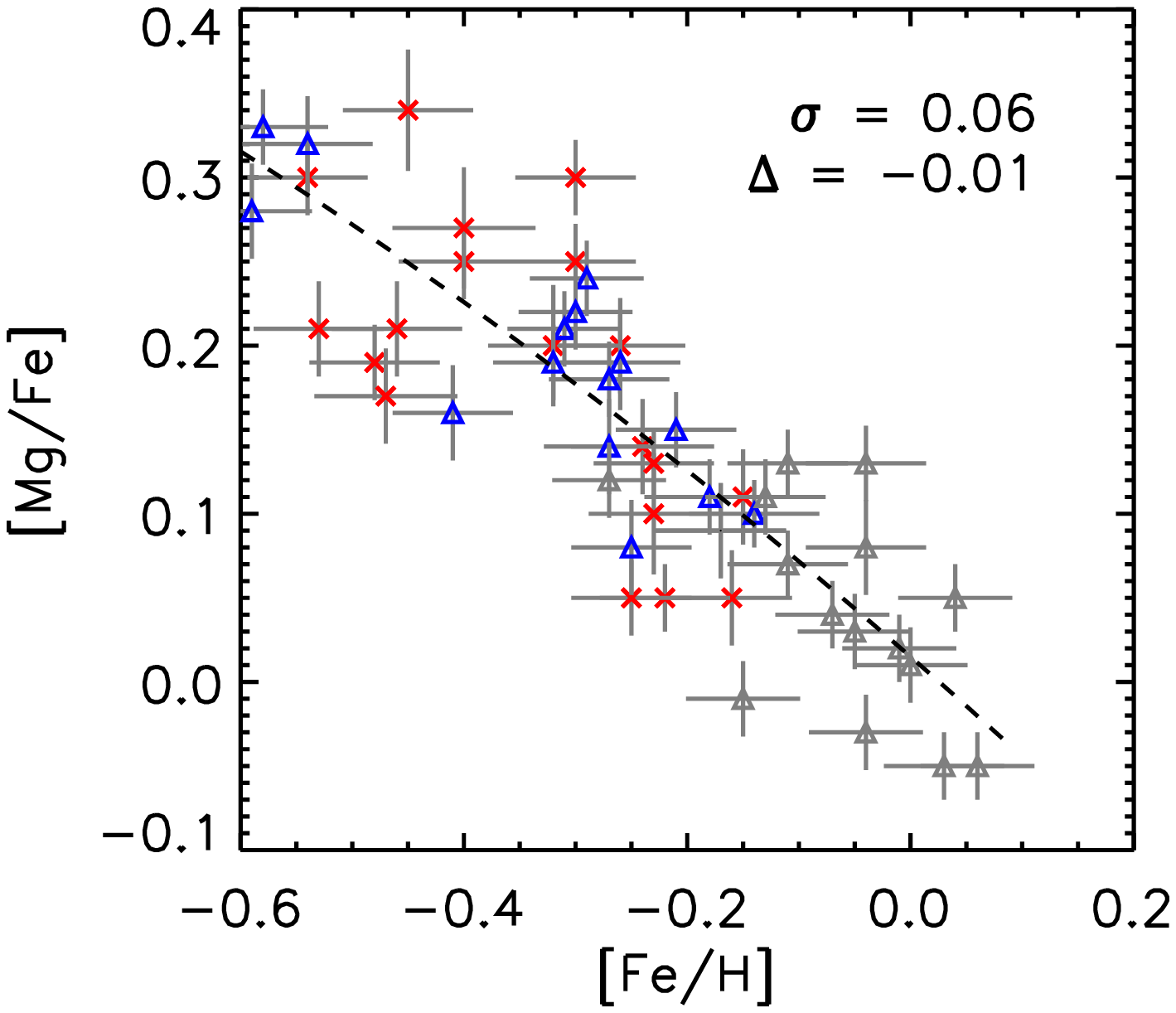}
\end{minipage}
\begin{minipage}{0.24\linewidth}
\includegraphics[width=\linewidth]{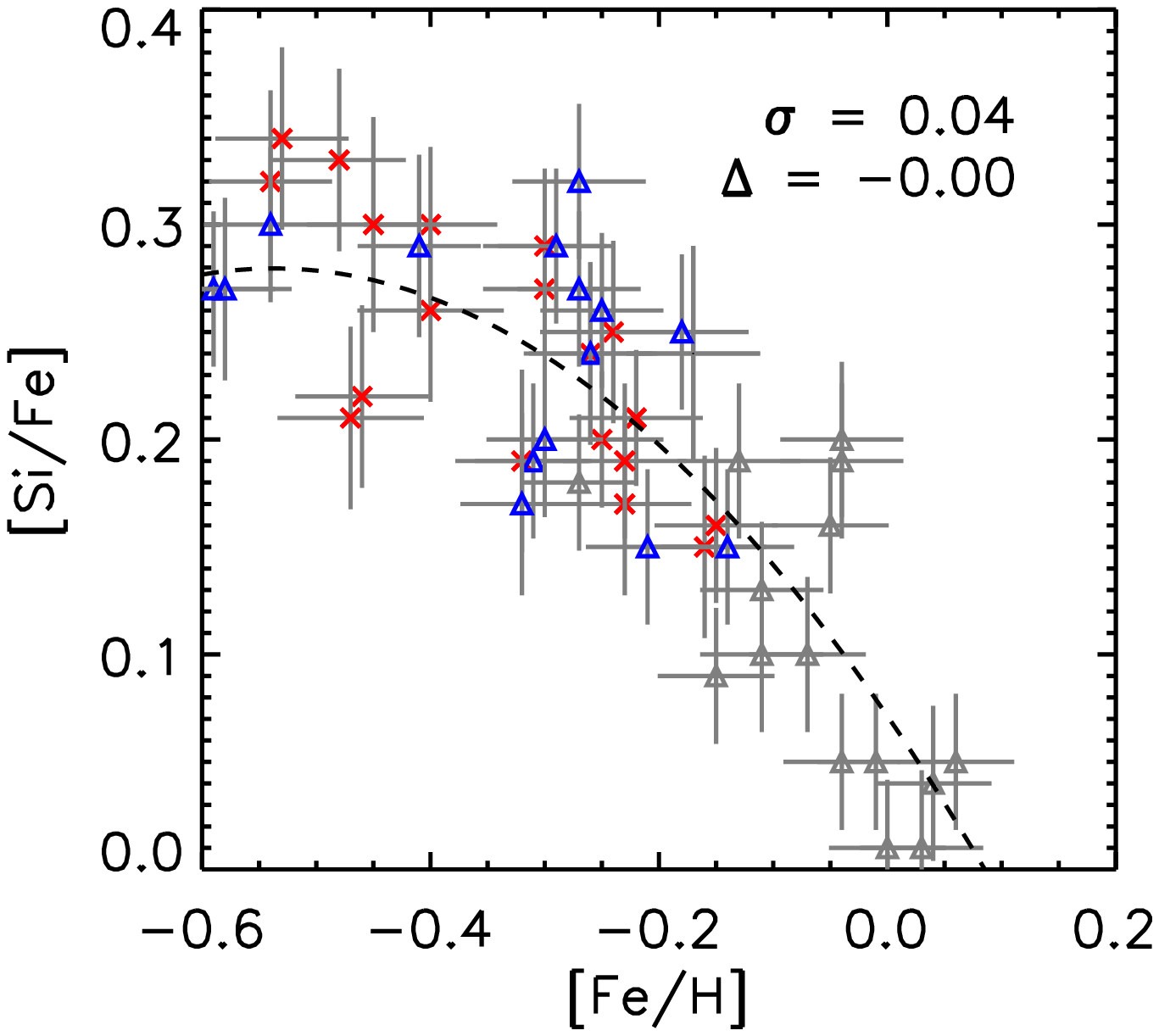}
\end{minipage}
\begin{minipage}{0.24\linewidth}
\includegraphics[width=\linewidth]{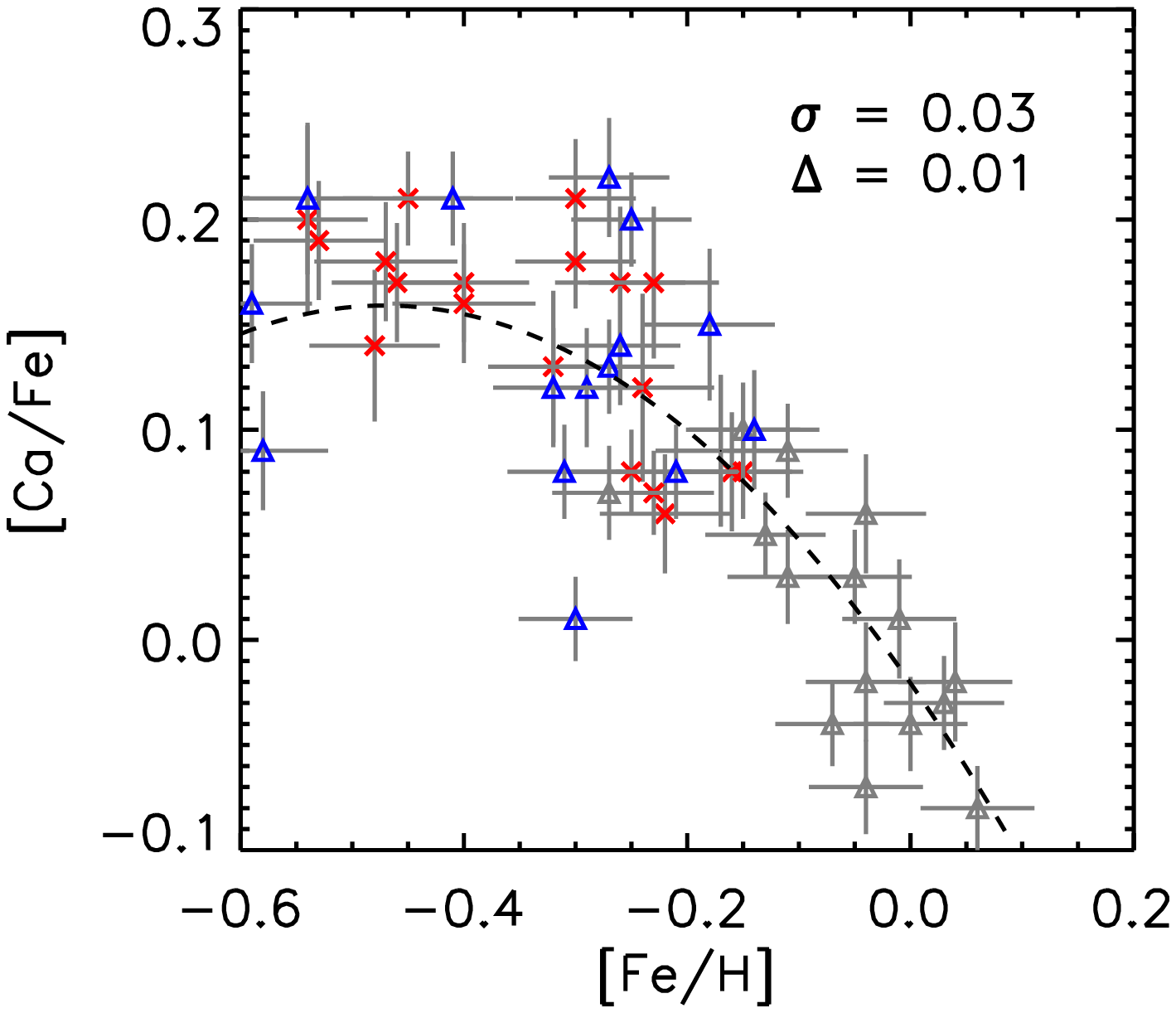}
\end{minipage}
\begin{minipage}{0.24\linewidth}
\includegraphics[width=\linewidth]{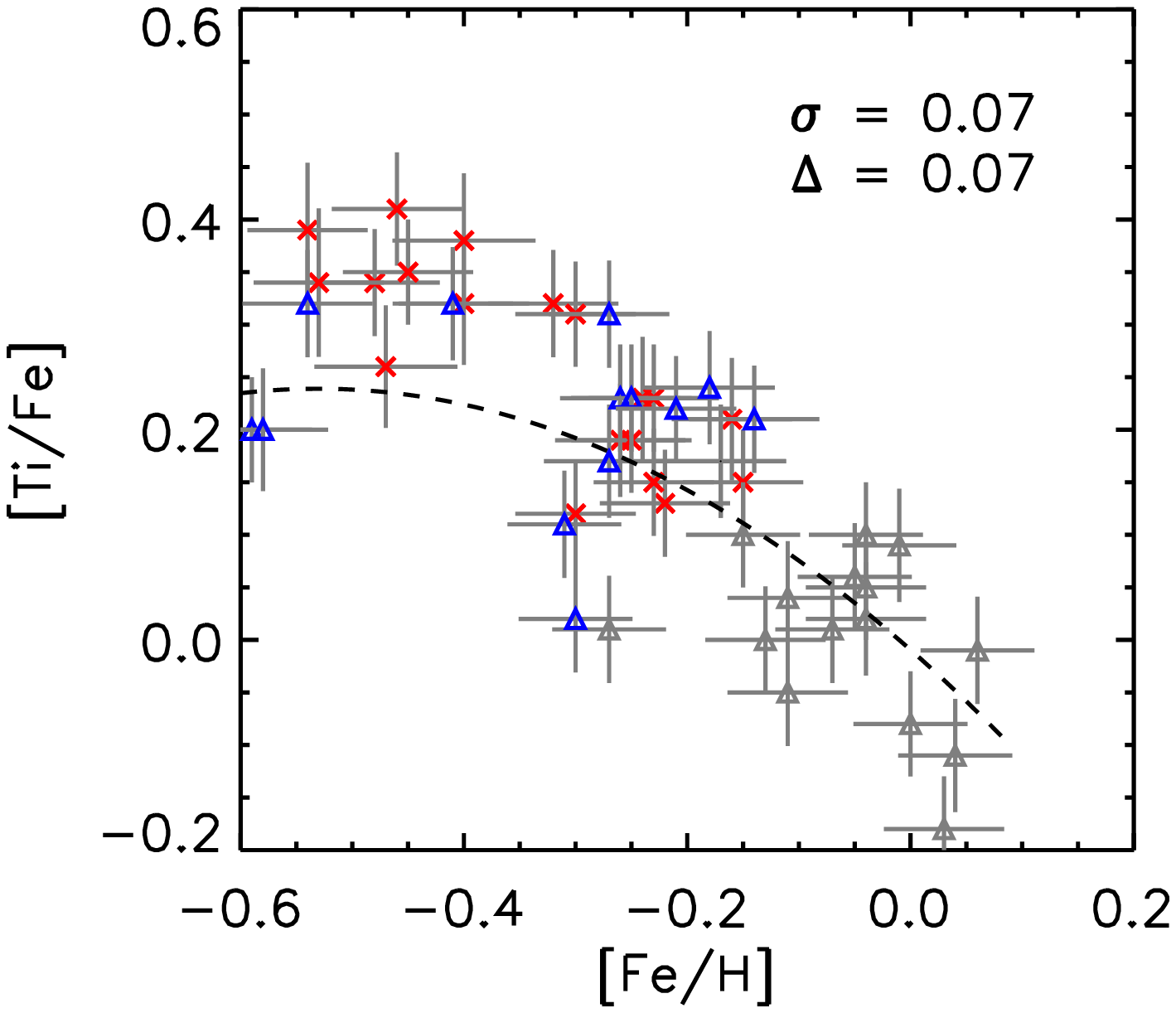}
\end{minipage}
\begin{minipage}{0.24\linewidth}
\includegraphics[width=\linewidth]{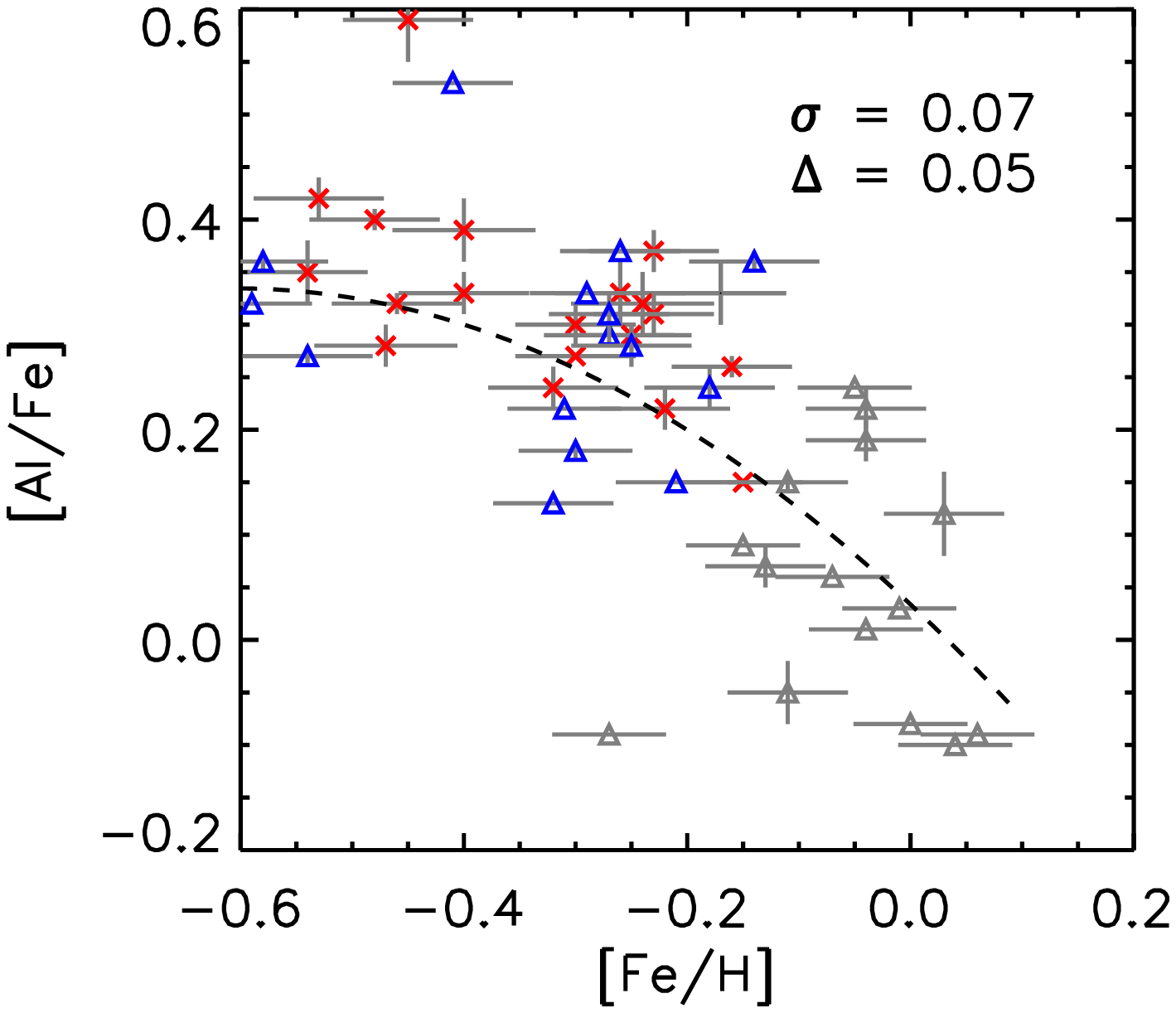}
\end{minipage}
\begin{minipage}{0.24\linewidth}
\includegraphics[width=\linewidth]{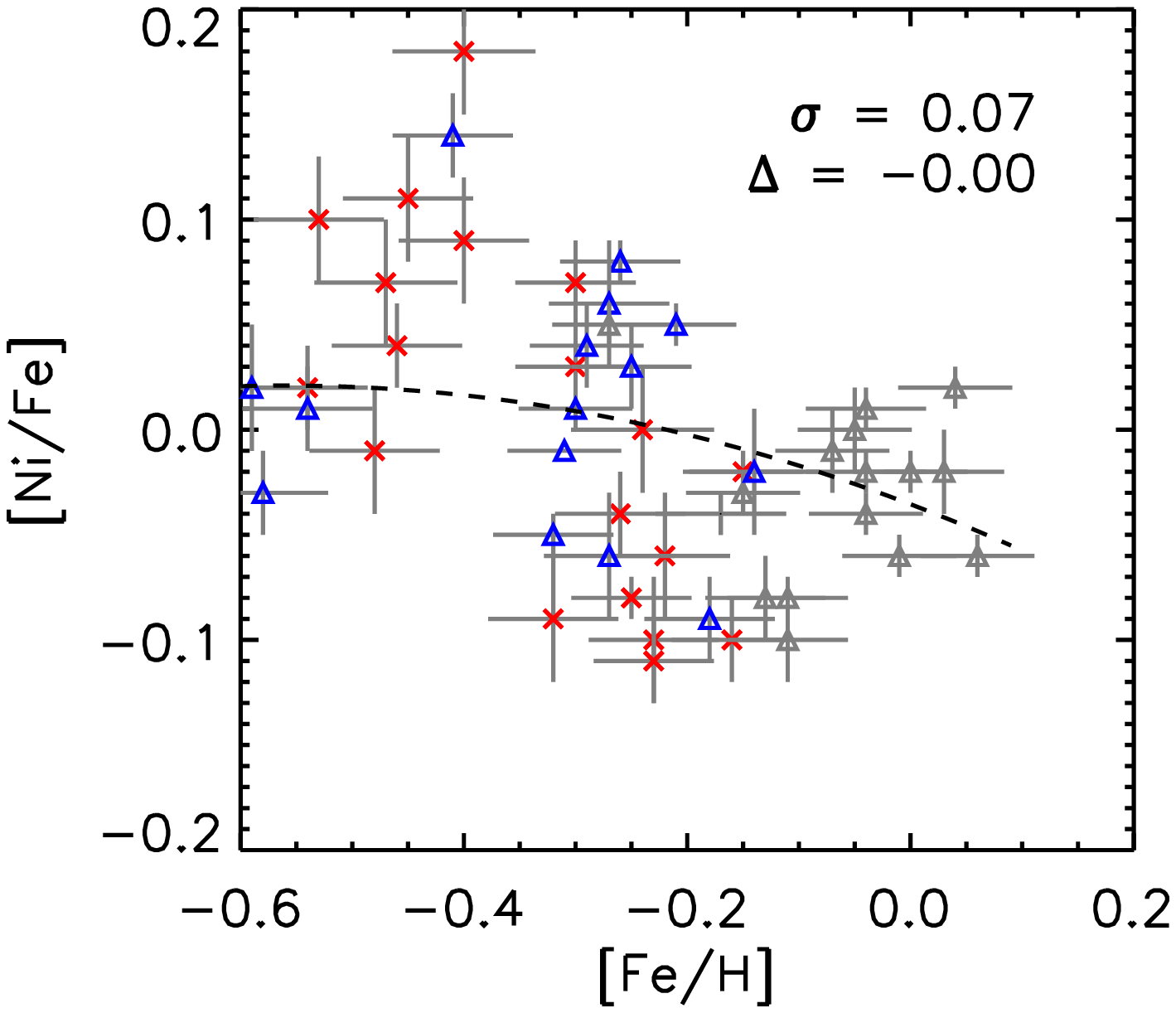}
\end{minipage}
\begin{minipage}{0.24\linewidth}
\includegraphics[width=\linewidth]{empty}
\end{minipage}
\begin{minipage}{0.24\linewidth}
\includegraphics[width=\linewidth]{empty}
\end{minipage}
\caption{[X/Fe] vs. [Fe/H] for the stars analysed in this work. The meaning of the symbols is indicated in the legend in the top right panel. The black dashed line is a second order polynomial fit through the combined $\alpha$-rich old and $\alpha$-normal set of stars and the number in the right top of each panel indicates the standard deviation of the values of the $\alpha$-rich young stars around the fit ($\sigma$) and the average deviation from the fit ($\Delta$).}
\label{elements}
\end{figure*}

We performed a differential abundance analysis relative to the Sun using the {\it abfind} driver of {\sc \bf MOOG} and adopting 1D model atmospheres \citep{caskur2003} based on the derived stellar parameters and the EWs of the absorption lines. In table \ref{chemical_abundances}, we list abundance results for individual stars averaged over all available lines of given species relative to solar abundances derived from the adopted $gf$-values. In Fig.~\ref{elements} we show the trends of [X/Fe] vs. [Fe/H] and in Tables~\ref{EWs0} - \ref{EWs6} we provide EW values for individual lines measured.

The computed synthetic spectra based on the derived stellar parameters were matched to the observed spectra by adjusting the abundances of the derived C, N and O abundances from combinations of vibration-rotation lines of CO and OH along with electronic transitions of CN. The adopted dissociation energies (D$_{\rm 0}$) in the spectrum synthesis of OH, CO and CN are 4.411 eV \citep{brooke2016}, 11.092 eV \citep{huber1979} and 7.724 eV \citep{sneden2014}, respectively. Following \citet{smith2013}, we measured the C, N and O abundances by fitting all three types of molecular lines consistently, i.e. we required molecular equilibrium: we first measured the oxygen abundance from the OH line, then derived the carbon abundance from CO in an iterative fashion. This process is repeated until the abundances of C and O via fits to the CO and OH lines yield consistent values. Then keeping these C and O abundances fixed, we derived the nitrogen abundance through the synthesis of multiple CN lines. The final adopted abundances of C, N and O are those values that provide self-consistent results from OH, CO and CN lines. We provide CNO abundance results for individual stars as well as the solar values in table \ref{cno_abundances} .

\begin{figure*}
\centering
\begin{minipage}{0.32\linewidth}
\includegraphics[width=\linewidth]{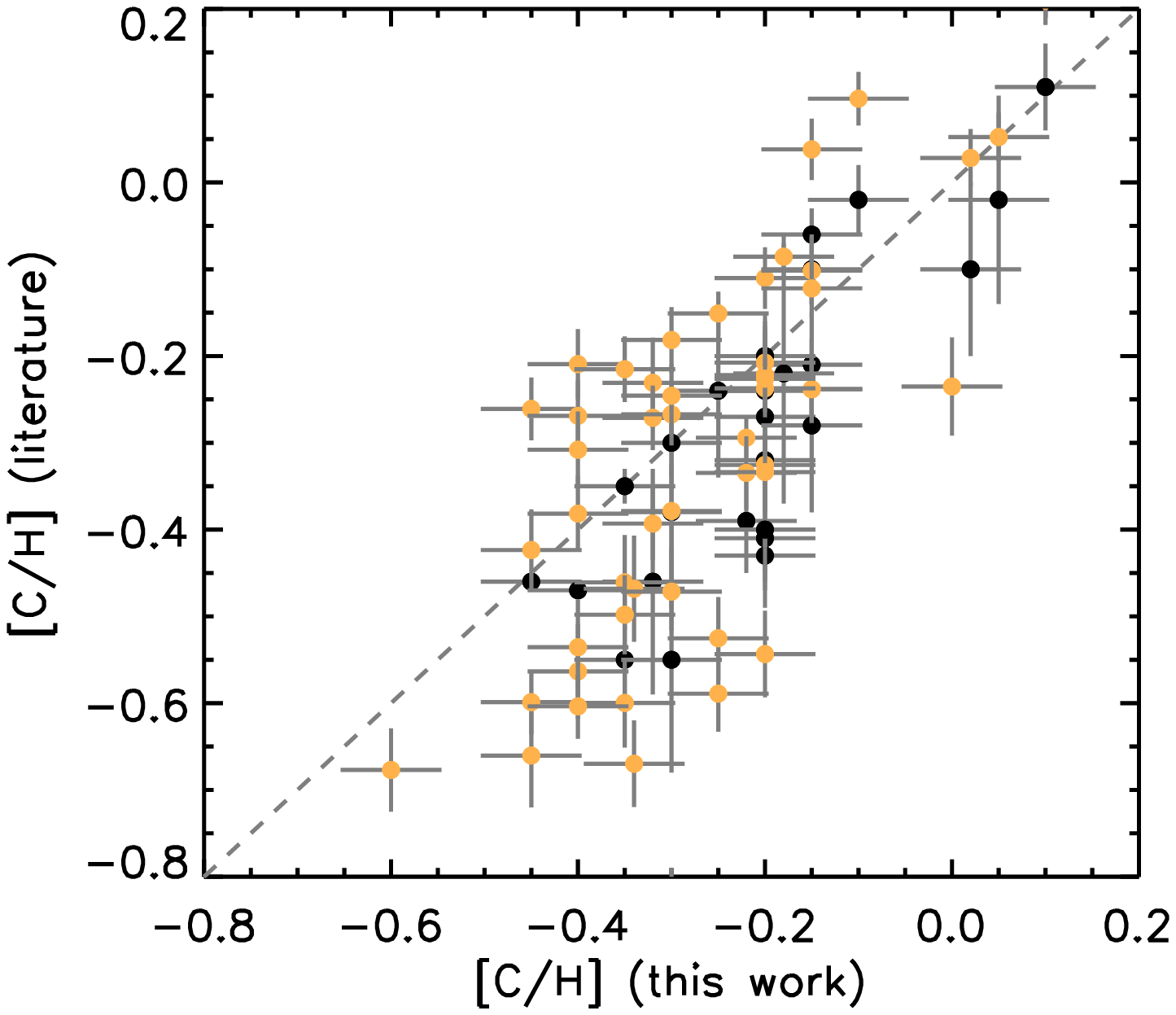}
\end{minipage}
\begin{minipage}{0.32\linewidth}
\includegraphics[width=\linewidth]{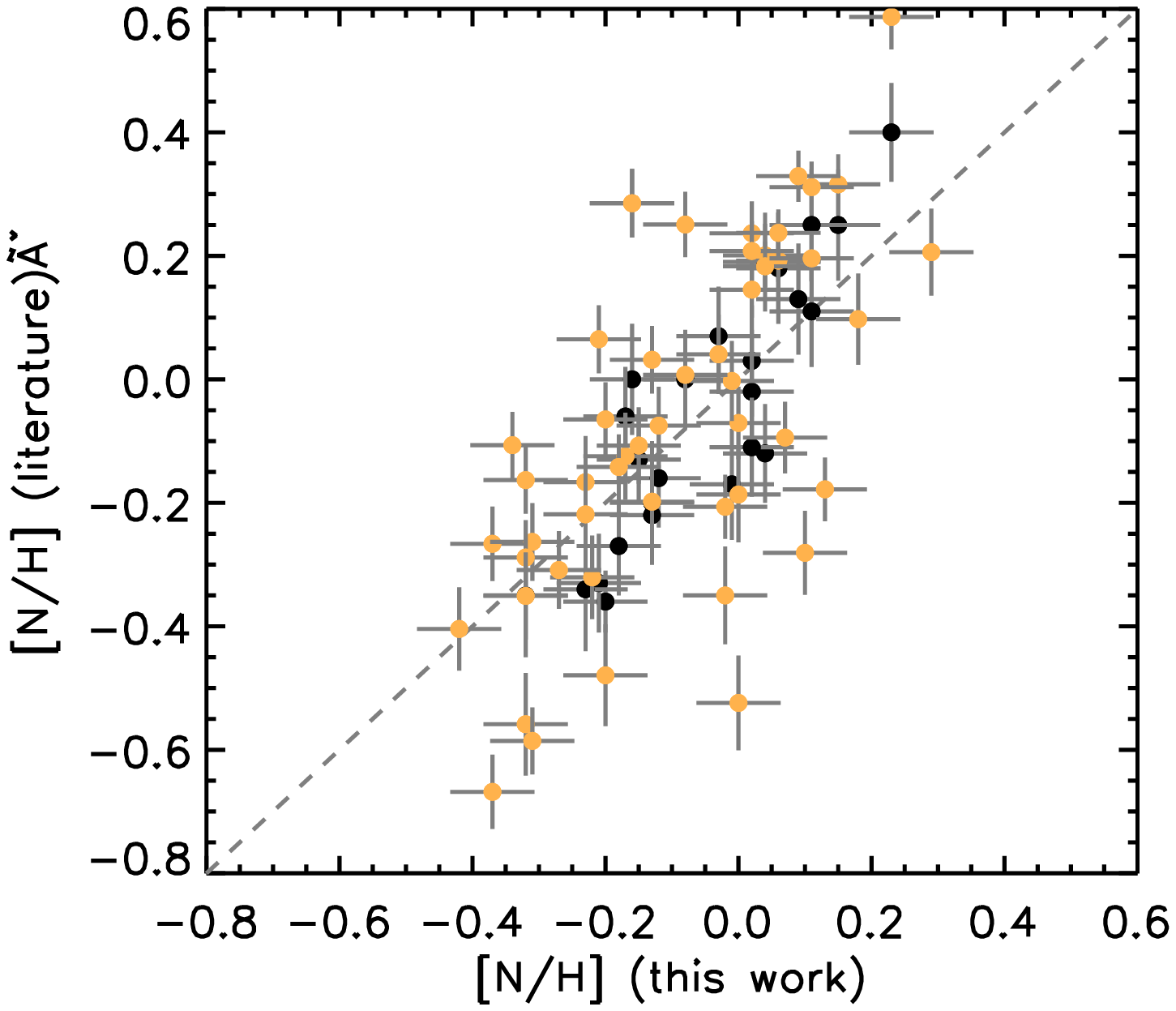}
\end{minipage}
\begin{minipage}{0.32\linewidth}
\includegraphics[width=\linewidth]{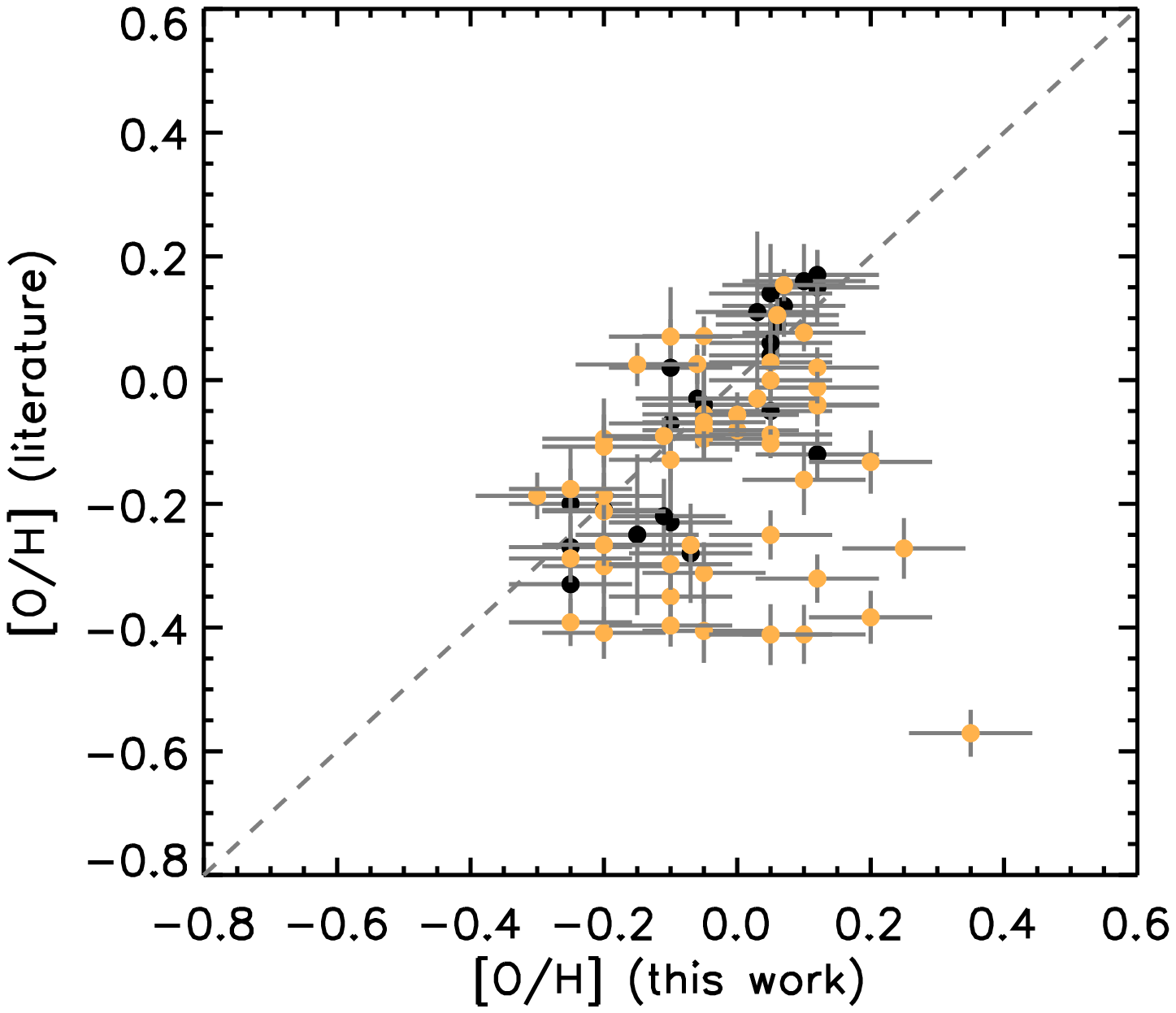}
\end{minipage}
\caption{Comparisons between [C/H] results (left),  [N/H] results (middle) and [O/H] results (right) obtained in this work and with the BACCHUS pipeline \citep[black dots,][]{hawkins2016} and with the APOGEE DR14 catalogue \citep[orange dots,][]{holtzman2018}. The gray dashed line indicates the one-to-one relation.}
\label{comp_cno}
\end{figure*}

\subsection{Uncertainties}
Stars in our sample are analysed identically and span small ranges in atmospheric parameters. Thus, systematic uncertainties affecting the abundances and abundance ratios [X/Fe] are expected to be consistent (and small) across the sample which spans a range in metallicity of $\sim$ $-$0.6 to $+$0.05 dex. 
Most of the elements analysed for chemical abundances are represented by more than one atomic or molecular line, and the internal consistency in the average abundance of an element is given by the $\pm$1$\sigma$ standard deviation about the mean abundance (later referred to as $\sigma_{1}$), which is typically less than 0.04 dex.
As the multiple lines of each element have varying line strengths and excitation potentials, the respective 1$\sigma$ uncertainties also provide an indication of how well the 1D LTE line analysis can reproduce all spectral lines in the observed spectra. 

Following the discussion in \cite{smith2013}, we evaluated the magnitude of uncertainties introduced in the measured chemical abundances by the uncertainties in stellar parameters ($T_{\rm eff}$, log~$g$, $\xi_{t}$) and metallicity by varying each parameter separately by an amount equal to its uncertainty, while keeping the other parameters unchanged. As the sample of red giants analysed in this paper spans a limited range in {\it T}$_{\rm eff}$, log~$g$ and $\xi_{t}$, the abundance sensitivity to stellar parameters is carried out for a representative star KIC\,6664950 with $T_{\rm eff}$~=~4800~K, log~$g$~=~2.4~dex and $\xi_{t}$~=~1.7~km\,s$^{-1}$. The incremental changes about the average abundance of an element caused by varying {\it T}$_{\rm eff}$, log~$g$, $\xi_{t}$ and the model metallicity by $\pm$ 50 K, 0.2 dex, 0.2 km s$^{-1}$ and 0.1 dex \citep[see for details of these choices][]{reddy2015}, respectively, with respect to the chosen model parameters are summarised in Table \ref{abu_sensitivity}. The square root of the quadratic sum of the mean values of all four contributors are presented in the column headed $\sigma_{2}$. The total error $\sigma_{\rm tot}$ in the measured elemental abundance is the square root of the quadratic sum of $\sigma_{1}$ and $\sigma_{2}$. Entries in Table \ref{abu_sensitivity} reveal that the observed dispersions in [El/Fe] for many of the elements are in the range 0.02$-$0.06 dex comparable to their respective $\pm$1$\sigma$ measurement uncertainties. Exceptions include the element oxygen whose uncertainty of 0.1 dex is dominated by the sensitivity of O abundance to the overall metallicity.

\subsection{Comparisons with the literature}
We compare our results with the ones obtained by \citet{hawkins2016} and the APOGEE collaboration as presented by \citet{holtzman2018}. The BACCHUS code \citep{hawkins2016} relies on the radiative transfer code Turbospectrum \citep{alvarez1998,plez2012} and the MARCS model atmosphere grid \citep{gustafsson2008}. The abundance are derived by comparing the observed spectrum with a set of convolved synthetic spectra characterised by different abundances. A $\chi^2$ diagnostic was used to obtain a robust abundance measurement. The ASPCAP pipeline \citep{garciaperez2016, holtzman2018} determines stellar parameters and abundances by finding the best match between observed spectra and a large grid of synthetic spectra. The synthetic spectral  grid is multi-dimensional including $T_{\rm eff}$, $\log(g)$, [M/H], [$\alpha$/M], [C/M], [N/M], and microturbulent velocity. In the $T_{\rm eff}$ range that we are interested in, ASPCAP uses a set of model atmospheres specifically generated for APOGEE: the APOGEE ATLAS9 models \citep{meszaros2012,zamora2015}, which are based on the ATLAS9 model atmosphere code from \citet{castelli2004}. These model atmospheres were constructed with the same C, N and $\alpha$ abundances, as well as [Fe/H], that were used to generate the synthetic spectrum at that grid point.

In this work we adopted $T_{\rm eff}$ and log~$g$ from \citet{hawkins2016} as starting values for the analysis of the set of 26 stars presented by \citet{jofre2016}. These values are the ones provided by APOGEE DR12. For the set of 25 stars that we newly selected in this work we used $T_{\rm eff}$ and log~$g$ from the APOGEE DR14 as starting values for the analysis. In Fig.~\ref{comp_fealpha} we show the comparisons between the metallicities, the $\alpha$ abundances as well as $T_{\rm eff}$ and log~$g$. For log~$g$ all comparisons are close to the one-to-one line. For $T_{\rm eff}$ there is a group of orange dots that are located below the one-to-one line. These are the stars selected by \citet{jofre2016} for which we used the \citet{hawkins2016}, i.e. APOGEE DR12 values of $T_{\rm eff}$ and log~$g$, and this offset is reminiscent of the offset in $T_{\rm eff}$ between APOGEE DR12 and DR14 values.

There exists a constant offset with small dispersion in [Fe/H] between the literature and values derived in our study. The typical mean difference between our and APOGEE DR14 metallicities is $+$0.09$\pm$0.08 dex (51 stars) and between our and \citet{hawkins2016} [Fe/H] estimates is $+$0.15$\pm$0.07 dex (26 stars), see also Table~\ref{offset}. Both dispersions about the mean differences are comparable to the line-to-line abundance dispersion observed for Fe lines. The comparable offsets of $+$0.15 dex and $+$0.09 dex in [Fe/H] between our study and the literature results provide an opportunity to assess the consistency of our abundance estimates.
For stars in common between the APOGEE DR14 and BACCHUS analyses, the typical mean difference in [Fe/H] is $-$0.12 dex with a dispersion of 0.05 dex, corresponding to the line-to-line abundance dispersion for Fe.
Therefore, the association of very similar offset (left panel in Figure \ref{comp_fealpha}) of our [Fe/H] values from \cite{hawkins2016} and the calibrated [M/H] from ASPCAP strengthens the assumption that our measured metallicities are self-consistent across the sample of 51 giants.

For [$\alpha$/Fe] we show a comparison in the central top panel of Fig.~\ref{comp_fealpha}. For both our work and for the BACCHUS pipeline the [$\alpha$/Fe] ratio is obtained by taking the average of [X/Fe] for the elements Mg, Si, Ca and Ti. For the APOGEE comparison we use the provided [$\alpha$/Fe] that results from the multi-dimensional grid matching (see above). Note that we do have stars with [$\alpha$/Fe] < 0.1 in the sample mostly originating from the \citet{jofre2016} selection. In Tables~\ref{stellarproperties}, \ref{chemical_abundances} and \ref{cno_abundances} we categorise stars with [$\alpha$/Fe] > 0.1 as $\alpha$-rich stars (21 young stars and 15 old stars) and stars with  [$\alpha$/Fe] $\leq$ 0.1 as $\alpha$-normal stars (15 stars).

We compare in Figure \ref{comp_cno} our LTE measured abundance ratios of [C/H], [N/H], [O/H] with results presented by \cite{hawkins2016} (black) for 26 giants in common between the studies and with results from the APOGEE DR14 catalogue \citep[orange,][]{holtzman2018} for 51 stars. We find general agreement with the BACCHUS and APOGEE DR14 results. However for the APOGEE results we find significantly more scatter. Because the [$\alpha$/Fe] measurement for APOGEE is derived by searching for the best match within a grid of synthetic spectra, it is most sensitive to the elements that have the most effect on the spectra. For the H-band, this element is O, instead of the elements used in the other [$\alpha$/Fe] calculations, which may account for part of the scatter. Additionally, we re-normalised the continuum in this work and used some additional features.

Summarising, Figs~\ref{comp_fealpha} and \ref{comp_cno} highlight satisfactory agreement of chemical abundances for all the elements despite some biases in [Fe/H]. As we perform a homogeneous analysis such biases will not impact on our final results.

\begin{table*}
\centering
\caption{Typical mean difference and dispersion (indicated after the $\pm$ symbol) between our and APOGEE DR14 \citep{pinsonneault2018} parameters (51 stars) as well as between our results and BACCHUS \citep{hawkins2016} parameters (26 stars) as well as for stars in common between the APOGEE DR14 and BACCHUS analyses.}
\label{offset}
\begin{tabular}{lccc}  
\hline
parameter & this work - APOGEE DR14 & this work - BACCHUS & BACCHUS - APOGEE DR14\\
\hline
$T_{\rm eff}$ [K] & $-38\pm58$ & $+0.4\pm1.8$ & $-75\pm58$\\
log~$g$ [dex] & $-0.001\pm0.046$ & $-0.001\pm0.019$ & $-0.007\pm0.009$\\
{[Fe/H]} & $+0.09\pm0.08$ & $+0.15\pm0.07$ & $-0.12\pm0.05$\\
{[$\alpha$/Fe]} & $-0.02\pm0.05$ & $-0.06\pm0.07$ & $+0.04\pm0.05$\\
{[$\alpha$/H]} & $+0.06\pm0.09$ & $+0.09\pm0.07$ & $-0.07\pm0.05$\\
{[C/H]} & $+0.04\pm0.15$ & $+0.07\pm0.10$ & $-0.10\pm0.06$\\
{[N/H]} & $-0.04\pm0.21$ & $-0.005\pm0.110$ & $-0.14\pm0.13$\\
{[O/H]} & $+0.12\pm0.21$ & $+0.01\pm0.09$ & $+0.02\pm0.13$\\
\hline
\end{tabular}
\end{table*}

\section{Chemical `peculiarities' in $\alpha$-rich young stars}
Abundance analyses by \citet{yong2016, matsuno2018} show that the majority of $\alpha$-rich young stars have abundances in line with thick disk stars and do not stand out apart from a few Li-rich stars, which are also seen among other samples of stars.
In Fig.~\ref{elements}, we show all individual abundance ratios as a function of [Fe/H] for our sample of stars. We investigated the dispersion of the abundance ratios of the $\alpha$-rich young
stars about a quadratic polynomial fit through the combined $\alpha$-rich old and alpha-normal sets of stars. For all elements except C, N and O the dispersion is small ($<$~0.075) and we attribute this to the uncertainties in [X/Fe] and [Fe/H]. The average deviation from the fit is typically close to zero except for [Al/Fe] and [Ti/Fe]. For C, N and O the dispersion is larger ($>$0.1) and additionally for [N/Fe] and [O/Fe] the values of the $\alpha$-rich young stars are predominantly higher than for the $\alpha$-rich old sample while towards lower [Fe/H] values there is a tendency for [C/Fe] of the $\alpha$-rich young stars to be lower than for the $\alpha$-rich old stars. (see values $\Delta$ for the average deviations from the fit).

As mentioned in the Introduction, the surface abundances of C and N are affected by first dredge-up, with higher mass stars dredging up material
with increased He and N and decreased C and Li \citep[e.g.][]{salaris2015}. The changes in N and C are a result of the burning in the interior, which is dominated by the CNO cycle. The offset in C and N for some of the $\alpha$-rich young stars is intriguing, as it could be related to higher masses for $\alpha$-rich young stars at first dredge-up.

\begin{figure*}
\centering
\begin{minipage}{0.49\linewidth}
\includegraphics[width=\linewidth]{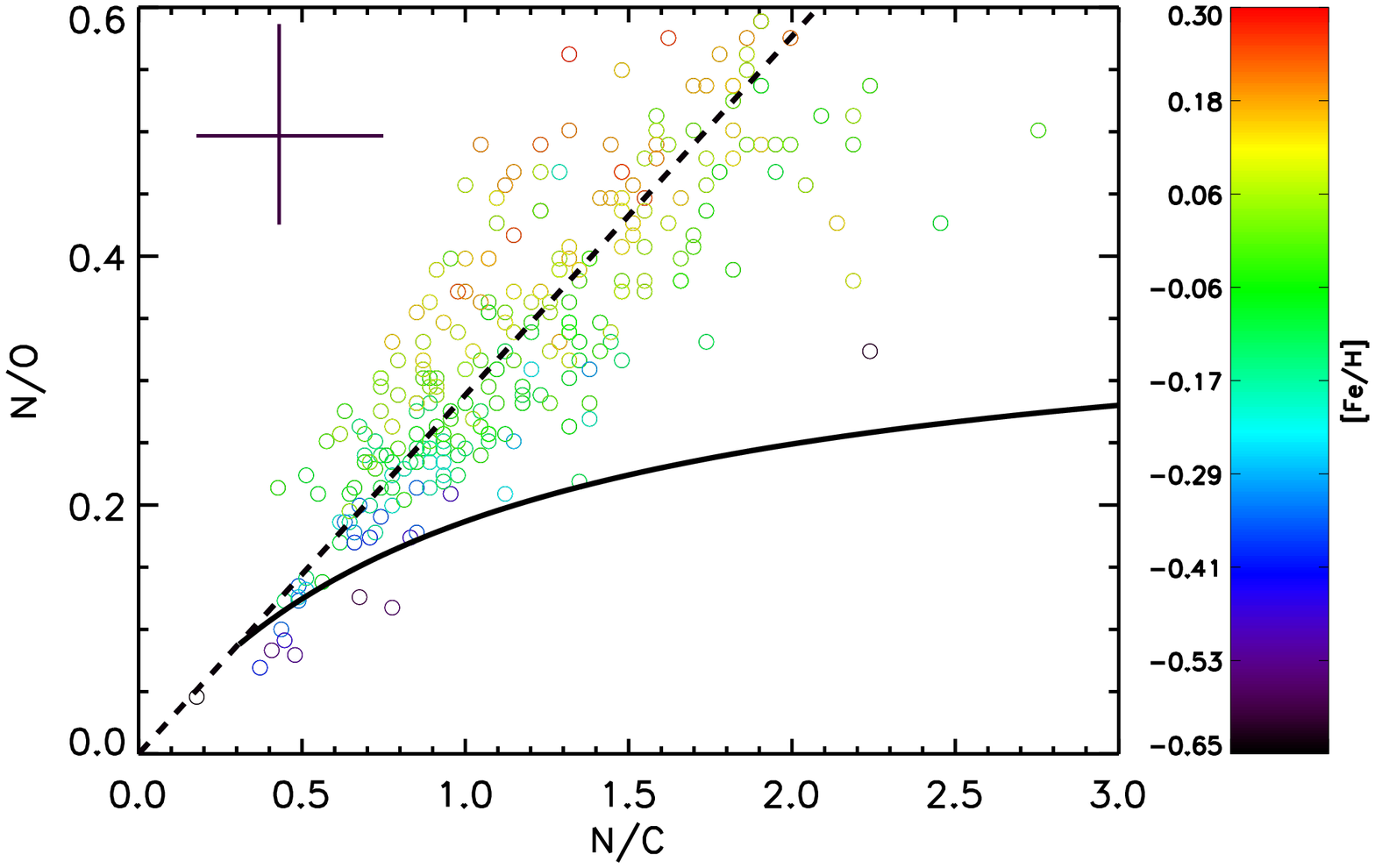}
\end{minipage}
\begin{minipage}{0.49\linewidth}
\includegraphics[width=\linewidth]{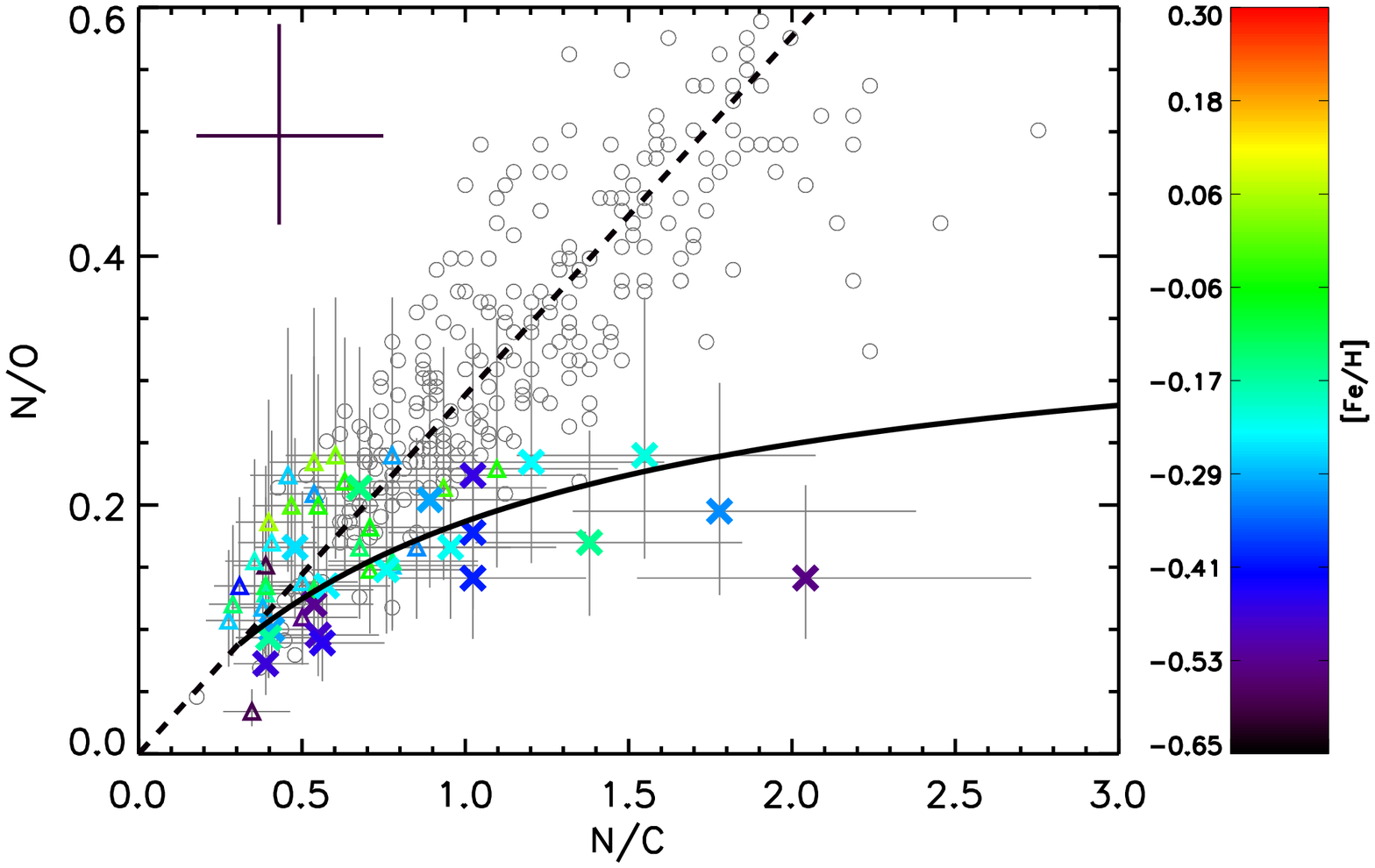}
\end{minipage}
\begin{minipage}{0.49\linewidth}
\includegraphics[width=\linewidth]{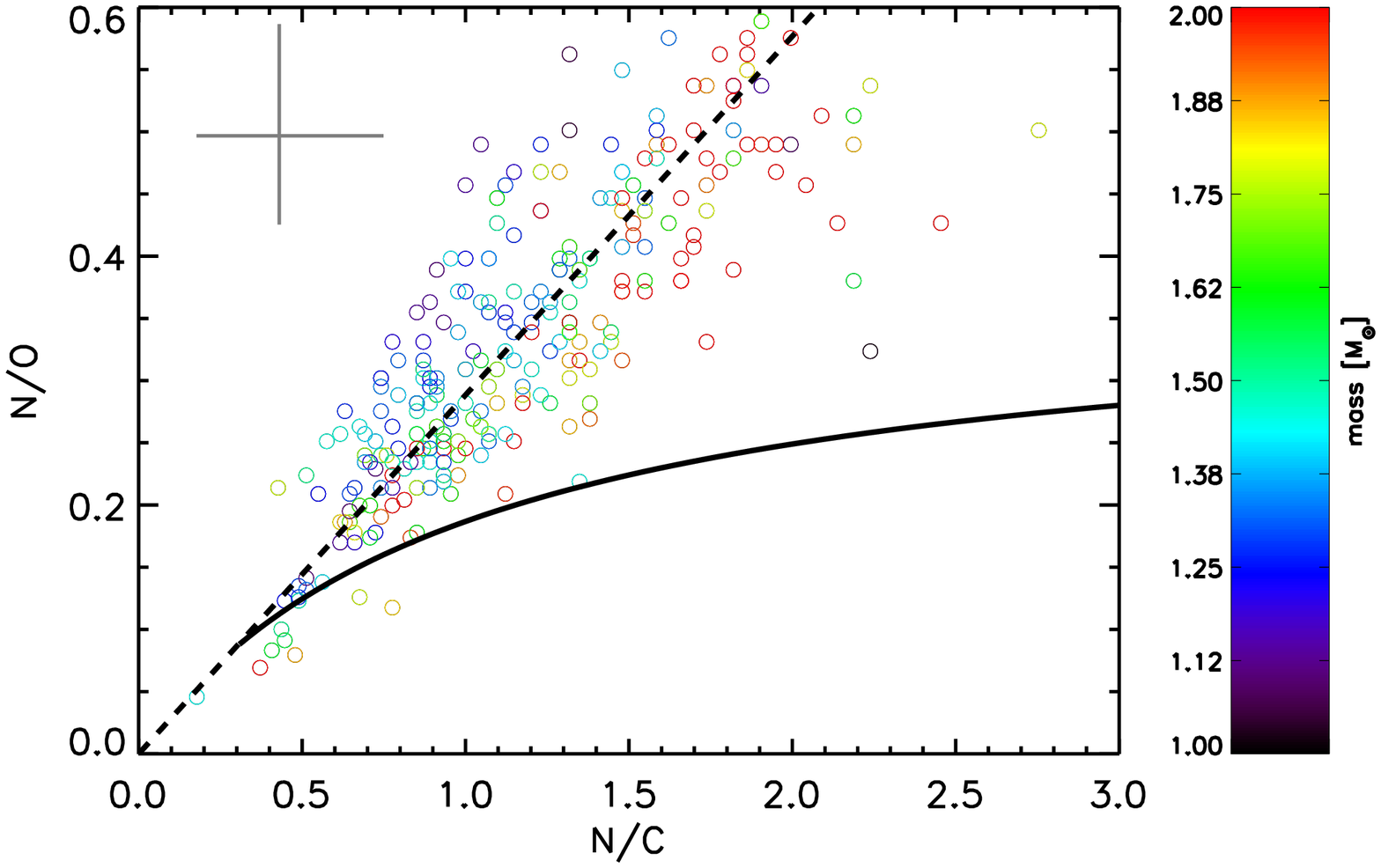}
\end{minipage}
\begin{minipage}{0.49\linewidth}
\includegraphics[width=\linewidth]{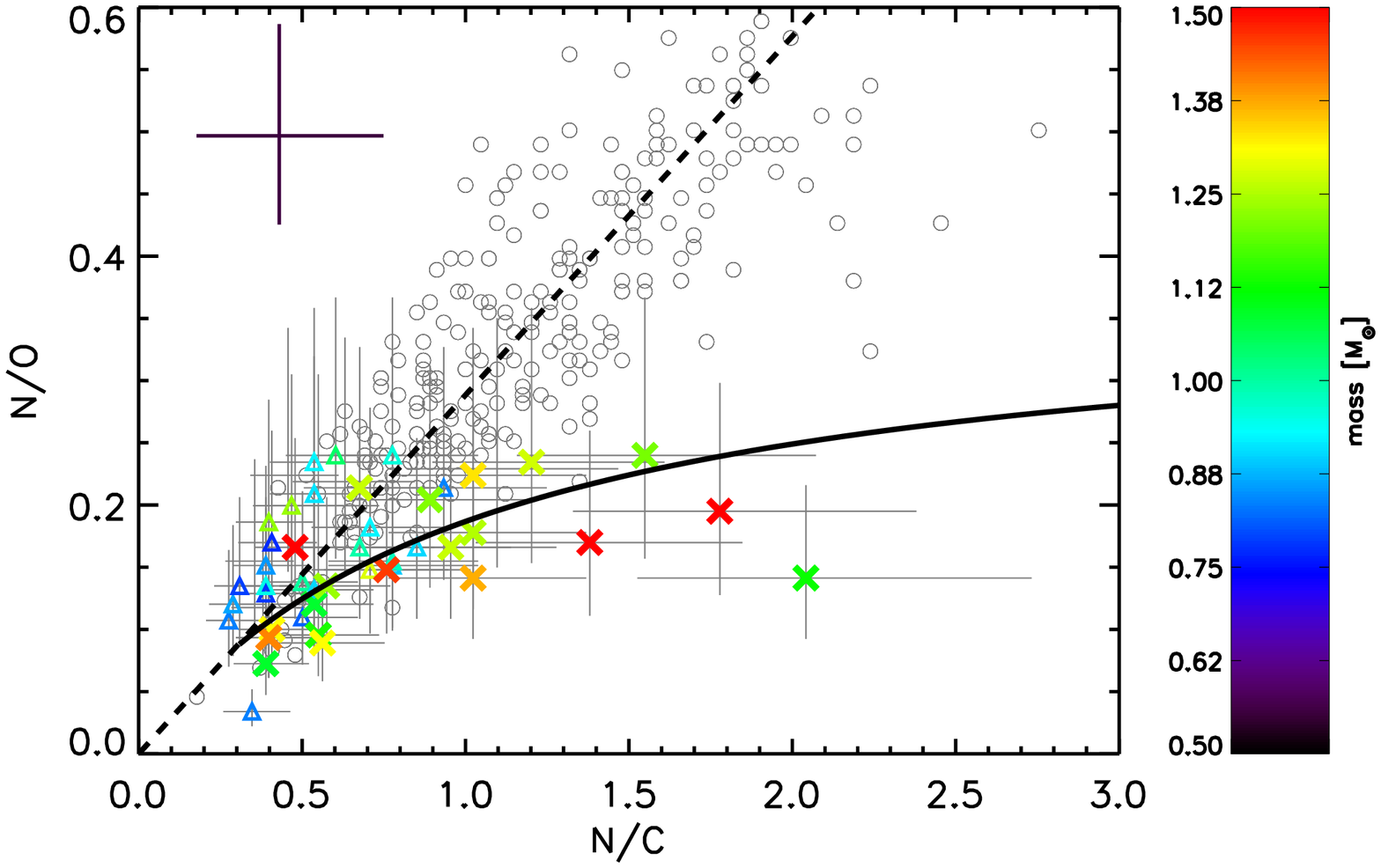}
\end{minipage}
\caption{N/O vs. N/C ratios for stars analysed by \citet{luck2007} (open coloured circles in the left panels) and the stars analysed in this work indicated with crosses ($\alpha$-rich young stars) and triangles ($\alpha$-rich old and $\alpha$-normal stars) in the right panels. In these right panels the stars analysed by \citet{luck2007} are indicated with gray circles. The colour code indicates the stellar metallicity (top) and mass (bottom). The masses of \citet{luck2007} and our work show an offset of about 0.5~M$_{\odot}$ and we have used two different colour scales to account for that. The black dashed and solid lines are toy-models explained in the text. The error bar in the upper left corner of each panel is representing the typical uncertainty in the \citet{luck2007} data.}
\label{CNOLuck}
\end{figure*}

We next look at the N/O versus N/C number density ratios to understand the impact of birth metallicity and mass of the stars on these ratios. We start with the metallicity. Typically speaking, in stars with higher metallicity we have a higher C, N and O initial abundance. Nitrogen is a secondary element, i.e. the yield increases with the increasing metallicity of the nucleosynthetic site \citep{talbot1974}. We have estimated the trend in N/C and N/O by assuming C/O to be constant and N is increasing linearly. This toy-model is overplotted over the \citet{luck2007} N/O vs. N/C data for field red giants in Fig.~\ref{CNOLuck} with the black dashed line. This is indeed consistent with the higher metallicity stars appearing at higher N/O and N/C.

The first dredge up will bring to the surface CN cycled material, where C has been converted to N. The bottom panel in Fig.~\ref{CNOLuck} shows that higher mass stars typically have higher N/C for a specific value of N/O. This can be explained by the fact that for higher mass stars the interior is hotter, which increases the N production. At the same time the dredge-up in higher mass star reaches to deeper (and thus hotter) layers. Therefore, in more massive stars more C is converted to N as compared to lower mass stars, and this is reflected in the surface abundances \citep[e.g.][]{karakas2014}. We model this mass effect for a single initial abundance of N/C=0.3 and N/O=0.09 by converting C to N and keeping O constant. This toy-model is shown by the solid line in Fig.~\ref{CNOLuck}.

The $\alpha$-rich young stars cover a large range in N/C, that is 9 out of 21 stars have N/C > 1, compared to 0 out of 15 and 1 out of 15 stars for the $\alpha$-rich old and $\alpha$-normal stars, respectively. If the stars are truly young and therefore always of higher mass, we expect them to appear at high N/C and indeed there are a number of them present at high N/C values. Additionally, we find some stars that are massive (i.e. green to red in the bottom panel of Fig.~\ref{CNOLuck}) that have relatively low values of N/C. So these are low-mass stars based on their chemistry, though they are now massive. We propose the following interpretation for this: these stars have the dredge-up features of low-mass stars. Their current high masses can be explained by a merger or mass-transfer scenario during or after dredge-up. \citet{izzard2018} indeed proposes that most stars with reasonable lifetimes to be observed will merge at that evolutionary stages. In fact some of the high mass high N/C stars could be the result of merger or mass transfer on the main sequence. Finally, we note that this would imply that the age determination as proposed by \citet{martig2016} based on C and N abundances for the $\alpha$-rich young stars that have merged or undergone mass-transfer during or after dredge-up, will be more accurate compared to methods that determine age as a function of mass.

To verify if all the stars can come from merger or mass transfer we need to consider the observed relative populations of high N/C and low N/C massive stars, compared to expectations from binary population synthesis models. Our pre-selected small sample may be sensitive to selection effects. The full set of APOGEE stars that overlaps with \textit{Kepler} data \citep{pinsonneault2018} and with the K2 data, will provide a more significant number of $\alpha$-rich young stars with which the proposed explanation can be confirmed (Johnson et al. in prep.).

\section*{Acknowledgements}
We thank A. B. S. Reddy for his contributions in the data analysis and preparation of this paper. The authors thank all members of the SAGE group at MPS and Marc Pinsonneault for useful discussions and input that improved the paper considerably. We would like to thank the anonymous referee for constructive suggestions to the paper that made it more clear and more helpful to the reader.
The research leading to the presented results has received funding from the European Research Council under the European Community's Seventh Framework Programme (FP7/2007-2013) / ERC grant agreement no 338251 (StellarAges) and the MERAC foundation. JAJ acknowledges funding from NASA grant NNX17AJOG and NSF grant AST-1211853.
NSO/Kitt Peak FTS Solar spectrum used in this paper was produced by NSF/NOAO.
This publication makes use of data products from the Wide-field Infrared Survey Explorer, which is a joint project of the University of California, Los Angeles, and the Jet Propulsion Laboratory/California Institute of Technology, funded by the National Aeronautics and Space Administration.




\bibliographystyle{mnras}
\bibliography{msbib} 



\appendix

\section{Some extra material}

\begin{table*}
\centering
\caption{Stellar parameters of $\alpha$-rich young and old red giants and $\alpha$-normal stars. We report the $T_{\rm eff}$ and log $g$ values as used in the current work, for which we assume 50~K and 0.05~dex uncertainties respectively. We provide $M$, $R$ and age (with + and $-$ uncertainties) as per \citet{pinsonneault2018}, where we reproduce the `Ageold' flag for stars with ages older than the age of the Universe.}
\label{stellarproperties}
\begin{tabular}{lccccc}   
\hline
\multicolumn{1}{c}{KIC} &\multicolumn{1}{c}{$T_{\rm eff}$ [K]} &\multicolumn{1}{c}{log $g$ (cgs)} &\multicolumn{1}{c}{mass [M$_{\odot}$]} &\multicolumn{1}{c}{radius [R$_{\odot}$]} & \multicolumn{1}{c}{age [Gyr]} \\
\hline
  \multicolumn{6}{c}{$\alpha$-rich (young)} \\
8172784 & 5000 & 3.10 & 1.72$\pm$0.06 & 6.85$\pm$0.02 & 1.40${+ 0.21}{-0.23}$\\
4136835 & 5000 & 2.85 & 1.52$\pm$0.06 & 8.03$\pm$0.02 & 1.79${+ 0.27}{-0.25}$\\
4143460 & 4750 & 2.50 & 1.62$\pm$0.18 & 11.70$\pm$0.08 & 1.65${+ 0.95}{-0.60}$\\
4143460 & 4750 & 2.50 & 1.62$\pm$0.18 & 11.70$\pm$0.08 & 1.65${+ 0.95}{-0.60}$\\
6664950 & 4800 & 2.40 & 1.49$\pm$0.13 & 11.29$\pm$0.05 & 1.91${+ 0.65}{-0.51}$\\
8539201 & 5000 & 2.50 & 1.57$\pm$0.17 & 12.36$\pm$0.07 & 1.58${+ 0.80}{-0.59}$\\
5795626 & 5000 & 2.50 & 1.30$\pm$0.13 & 10.67$\pm$0.05 & 2.40${+ 0.83}{-0.76}$\\
10096113 & 4950 & 2.40 & 1.40$\pm$0.12 & 11.35$\pm$0.05 & 2.04${+ 0.64}{-0.58}$\\
3973813 & 4600 & 2.30 & 1.50$\pm$0.06 & 14.03$\pm$0.02 & 2.23${+ 0.40}{-0.39}$\\
1163359 & 4575 & 2.20 & 1.45$\pm$0.07 & 15.30$\pm$0.03 & 2.38${+ 0.45}{-0.43}$\\
12300740 & 4800 & 2.40 & 1.33$\pm$0.06 & 11.87$\pm$0.02 & 2.76${+ 0.44}{-0.40}$\\
5956977 & 4800 & 2.60 & 1.40$\pm$0.06 & 9.00$\pm$0.02 & 2.64${+ 0.39}{-0.38}$\\
11717920 & 4700 & 2.20 & 1.37$\pm$0.08 & 14.37$\pm$0.03 & 2.54${+ 0.62}{-0.57}$\\
6837256 & 4750 & 2.50 & 1.34$\pm$0.06 & 11.29$\pm$0.02 & 2.73${+ 0.44}{-0.43}$\\
9821622 & 4739 & 2.71 & 1.49$\pm$0.06 & 8.91$\pm$0.02 & 2.25${+ 0.34}{-0.33}$\\
4143460 & 4750 & 2.50 & 1.62$\pm$0.18 & 11.70$\pm$0.08 & 1.65${+ 0.95}{-0.60}$\\
4143460 & 4750 & 2.50 & 1.62$\pm$0.18 & 11.70$\pm$0.08 & 1.65${+ 0.95}{-0.60}$\\
11394905 & 4785 & 2.50 & 1.36$\pm$0.10 & 10.88$\pm$0.04 & 2.32${+ 0.52}{-0.43}$\\
11823838 & 4812 & 2.53 & 1.65$\pm$0.12 & 11.46$\pm$0.05 & 1.51${+ 0.39}{-0.32}$\\
5512910 & 4849 & 2.51 & 1.45$\pm$0.13 & 11.23$\pm$0.06 & 2.05${+ 0.77}{-0.61}$\\
10525475 & 4676 & 2.40 & 1.46$\pm$0.12 & 11.24$\pm$0.05 & 2.26${+ 0.71}{-0.63}$\\
  \multicolumn{6}{c}{$\alpha$-rich (old)} \\
3849996 & 4830 & 2.40 & 0.87$\pm$0.11 & 10.18$\pm$0.04 & 7.05${+ 1.16}{-1.18}$\\
7337994 & 4790 & 2.40 & 0.88$\pm$0.12 & 10.18$\pm$0.05 & 7.05${+ 1.66}{-1.66}$\\
2973894 & 4935 & 2.30 & 0.85$\pm$0.18 & 10.06$\pm$0.08 & 6.84${+ 3.23}{-2.80}$\\
4446181 & 4830 & 2.40 & 0.88$\pm$0.12 & 9.86$\pm$0.05 & 7.03${+ 1.72}{-1.56}$\\
4751953 & 4750 & 2.30 & 0.79$\pm$0.10 & 11.05$\pm$0.04 & 8.95${+ 1.14}{-1.12}$\\
4661299 & 4900 & 2.35 & 0.82$\pm$0.19 & 10.10$\pm$0.08 & 7.98${+ 4.18}{-3.74}$\\
4548530 & 4685 & 2.30 & 0.82$\pm$0.12 & 11.95$\pm$0.06 & Ageold\\
10095427 & 4625 & 2.60 & 0.98$\pm$0.06 & 7.68$\pm$0.02 & 9.27${+ 1.60}{-1.52}$\\
4480358 & 4500 & 2.00 & 1.01$\pm$0.14 & 19.60$\pm$0.06 & 6.28${+ 4.12}{-3.45}$\\
3936823 & 4500 & 2.10 & 0.83$\pm$0.09 & 14.13$\pm$0.04 & 13.71${+ 3.71}{-3.95}$\\
10586902 & 4606 & 2.53 & 1.06$\pm$0.06 & 9.00$\pm$0.02 & 7.26${+ 1.15}{-1.11}$\\
9157260 & 4724 & 3.28 & 1.01$\pm$0.07 & 3.80$\pm$0.03 & 9.98${+ 3.02}{-2.85}$\\
4844527 & 4774 & 2.38 & 0.95$\pm$0.12 & 10.12$\pm$0.05 & 6.05${+ 1.50}{-1.51}$\\
10463137 & 4819 & 2.38 & 0.93$\pm$0.12 & 10.26$\pm$0.05 & 6.47${+ 1.40}{-1.47}$\\
11870991 & 4792 & 2.39 & 0.97$\pm$0.11 & 10.21$\pm$0.04 & 5.74${+ 1.13}{-1.13}$\\
  \multicolumn{6}{c}{$\alpha$-normal} \\
4350501 & 4744 & 3.06 & 1.51$\pm$0.09 & 6.05$\pm$0.04 & 2.47${+ 0.85}{-0.78}$\\
9269081 & 4729 & 2.30 & 1.77$\pm$0.16 & 15.52$\pm$0.06 & 1.36${+ 0.63}{-0.34}$\\
11445818 & 4669 & 2.46& ... & ... & ... \\
3833399 & 4601 & 2.47 & 1.45$\pm$0.14 & 11.57$\pm$0.06 & 2.58${+ 1.02}{-0.89}$\\
9761625 & 4435 & 1.85 & 1.30$\pm$0.07 & 22.24$\pm$0.03 & 3.86${+ 0.86}{-0.82}$\\
3455760 & 4572 & 2.57 & 1.41$\pm$0.06 & 10.06$\pm$0.02 & 3.40${+ 0.50}{-0.49}$\\
9002884 & 4229 & 1.55 & 1.01$\pm$0.13 & 27.89$\pm$0.06 & 8.59${+ 5.63}{-4.47}$\\
2142095 & 4756 & 3.40 & 1.15$\pm$0.14 & 11.50$\pm$0.06 & 3.77${+ 1.35}{-1.21}$\\
8611114 & 4695 & 2.43 & 1.01$\pm$0.11 & 10.07$\pm$0.04 & 5.77${+ 1.10}{-1.04}$\\
7594865 & 4669 & 2.39 & 1.20$\pm$0.16 & 11.37$\pm$0.07 & 3.70${+ 1.76}{-1.48}$\\
1432587 & 4312 & 1.64 & 0.86$\pm$0.23 & 23.43$\pm$0.09 & Ageold\\
3658136 & 4411 & 1.78 & 0.91$\pm$0.16 & 20.02$\pm$0.07 & Ageold\\
10880958 & 4644 & 2.42 & 1.12$\pm$0.10 & 10.68$\pm$0.03 & 4.70${+ 0.67}{-0.68}$\\
9143924 & 4446 & 1.87 & 0.83$\pm$0.21 & 17.83$\pm$0.09 & Ageold\\
9605294 & 4602 & 2.38 & 1.01$\pm$0.13 & 10.57$\pm$0.05 & 6.40${+ 1.62}{-1.58}$\\
\hline
\end{tabular}
\end{table*}

\begin{table*}
\centering
\caption{Chemical composition of $\alpha$-rich young and old red giants and $\alpha$-normal stars analysed in this paper along with reference [X/H]-ratios derived for the Sun. }
\label{chemical_abundances}
\begin{tabular}{lcccccccc}   \hline
  Sun ([X/H]) & 7.47$\pm$0.03  &  6.18$\pm$0.04 & 7.46$\pm$0.02 &  6.35$\pm$0.02  & 7.45$\pm$0.03  & 6.25$\pm$0.03 & 4.93$\pm$0.02  & \\ \hline
\multicolumn{1}{c}{KIC} &\multicolumn{1}{c}{[Fe/H]} &\multicolumn{1}{c}{[Ni/Fe] } & \multicolumn{1}{c}{[Mg/Fe]} & \multicolumn{1}{c}{[Al/Fe]} & \multicolumn{1}{c}{[Si/Fe]} & \multicolumn{1}{c}{[Ca/Fe]} & \multicolumn{1}{c}{[Ti/Fe]}  & \multicolumn{1}{c}{[$\alpha$/Fe]}\\ \hline
  \multicolumn{9}{c}{$\alpha$-rich (young)} \\
 4169517 & $-0.26\pm$0.03 & $-0.04\pm$0.02 & +0.20$\pm$0.02 & +0.33$\pm$0.03 & +0.24$\pm$0.03 & +0.17$\pm$0.03 & +0.19$\pm$0.02  & +0.20$\pm$0.04 \\
 3946701 & $-0.32\pm$0.03 & $-0.09\pm$0.03 & +0.20$\pm$0.03 & +0.24$\pm$0.02 & +0.19$\pm$0.03 & +0.13$\pm$0.03 & +0.32$\pm$0.01 & +0.21$\pm$0.04  \\
 8172784 & $-0.16\pm$0.02 & $-0.10\pm$0.02 & +0.05$\pm$0.02 & +0.26$\pm$0.01 & +0.15$\pm$0.03 & +0.08$\pm$0.02 & +0.21$\pm$0.03  & +0.12$\pm$0.04 \\
 4136835 & $-0.40\pm$0.03 & +0.09$\pm$0.03 & +0.25$\pm$0.01 & +0.33$\pm$0.02 & +0.30$\pm$0.02 & +0.17$\pm$0.02 & +0.32$\pm$0.03  & +0.26$\pm$0.04  \\
 4143460 & $-0.17\pm$0.03 & $-0.04\pm$0.01 & +0.09$\pm$0.02 & +0.33$\pm$0.03 & +0.24$\pm$0.04 & +0.09$\pm$0.03 & +0.17$\pm$0.02  & +0.15$\pm$0.04 \\
 6664950 & $-0.24\pm$0.04 & +0.00$\pm$0.03 & +0.14$\pm$0.02 & +0.32$\pm$0.03 & +0.25$\pm$0.03 & +0.12$\pm$0.04 & +0.23$\pm$0.03  & +0.19$\pm$0.04 \\
 8539201 & $-0.46\pm$0.03 & +0.04$\pm$0.02 & +0.21$\pm$0.02 & +0.32$\pm$0.01 & +0.22$\pm$0.03 & +0.17$\pm$0.02 & +0.41$\pm$0.02 & +0.25$\pm$0.04  \\
 5795626 & $-0.48\pm$0.03 & $-0.01\pm$0.03 & +0.19$\pm$0.01 & +0.40$\pm$0.01 & +0.33$\pm$0.03 & +0.14$\pm$0.03 & +0.34$\pm$0.01 & +0.25$\pm$0.04 \\
 10096113 & $-0.45\pm$0.03 & +0.11$\pm$0.03 & +0.34$\pm$0.03 & +0.59$\pm$0.04 & +0.30$\pm$0.04 & +0.21$\pm$0.01 & +0.35$\pm$0.00 & +0.30$\pm$0.04 \\
 3973813 & $-0.23\pm$0.03 & $-0.10\pm$0.03 & +0.10$\pm$0.03 & +0.37$\pm$0.02 & +0.17$\pm$0.03 & +0.17$\pm$0.03 & +0.23$\pm$0.01  & +0.17$\pm$0.04 \\
 1163359 & $-0.22\pm$0.03 & $-0.06\pm$0.03 & +0.05$\pm$0.00 & +0.22$\pm$0.02 & +0.21$\pm$0.01 & +0.06$\pm$0.02 & +0.13$\pm$0.01 & +0.11$\pm$0.03  \\
12300740 & $-0.54\pm$0.02 & +0.02$\pm$0.02 & +0.30$\pm$0.01 & +0.35$\pm$0.03 & +0.32$\pm$0.03 & +0.20$\pm$0.04 & +0.39$\pm$0.04 & +0.30$\pm$0.04  \\
 5956977 & $-0.40\pm$0.04 & +0.18$\pm$0.03 & +0.27$\pm$0.03 & +0.39$\pm$0.03 & +0.26$\pm$0.03 & +0.16$\pm$0.02 & +0.38$\pm$0.04 & +0.27$\pm$0.04  \\
11717920 & $-0.53\pm$0.03 & +0.10$\pm$0.03 & +0.21$\pm$0.02 & +0.42$\pm$0.02 & +0.34$\pm$0.03 & +0.19$\pm$0.02 & +0.34$\pm$0.05 & +0.27$\pm$0.04  \\
 6837256 & $-0.47\pm$0.04 & +0.07$\pm$0.03 & +0.17$\pm$0.02 & +0.28$\pm$0.02 & +0.21$\pm$0.03 & +0.18$\pm$0.02 & +0.26$\pm$0.03 & +0.21$\pm$0.04 \\
 9821622 &  $-0.30\pm$0.02 & +0.07$\pm$0.02 & +0.30$\pm$0.01 & +0.30$\pm$0.02 &  +0.27$\pm$0.02 &  +0.21$\pm$0.02 &  +0.31$\pm$0.00 & +0.27$\pm$0.03\\
 4143460 & $-0.17\pm$0.03 & $-0.04\pm$0.01   & +0.09$\pm$0.02 & +0.33$\pm$0.03 & +0.24$\pm$0.04 & +0.09 $\pm$0.03 & +0.17$\pm$0.02 & +0.15$\pm$0.04 \\ 
11394905 &  $-0.30\pm$0.02  &   +0.03$\pm$0.02 &   +0.25$\pm$0.01 &  +0.27$\pm$0.00 &  +0.29$\pm$0.02 &  +0.18$\pm$0.01 &  +0.12$\pm$0.01 & +0.21$\pm$0.03\\
11823838 &  $-0.23\pm$0.02  &   $-0.11\pm$0.02  &  +0.13$\pm$0.00 &  +0.31$\pm$0.02 &  +0.19$\pm$0.02 &  +0.07$\pm$0.00 &  +0.15$\pm$0.01 & +0.13$\pm$0.03 \\
 5512910 &  $-0.25\pm$0.02  &   $-0.08\pm$0.01  &  +0.05$\pm$0.01 &  +0.29$\pm$0.01 &  +0.20$\pm$0.01 &  +0.08$\pm$0.00 &  +0.19$\pm$0.00 & +0.13$\pm$0.03\\
 10525475 &  $-0.15\pm$0.02  &   $-0.02\pm$0.01 & +0.11$\pm$0.02 &  +0.15$\pm$0.00 &  +0.16$\pm$0.02 &  +0.08$\pm$0.01 &  +0.15$\pm$0.00 & +0.12$\pm$0.03 \\

  \multicolumn{9}{c}{$\alpha$-rich (old)} \\
 3849996 & $-0.27\pm$0.03 & $-0.06\pm$0.03 & +0.14$\pm$0.02 & +0.29$\pm$0.01 & +0.32$\pm$0.02 & +0.13$\pm$0.01 & +0.17$\pm$0.02 & +0.19$\pm$0.04 \\
 7337994 & $-0.14\pm$0.03 & $-0.02\pm$0.03 & +0.10$\pm$0.00 & +0.36$\pm$0.01 & +0.15$\pm$0.02 & +0.10$\pm$0.02 & +0.21$\pm$0.01 & +0.14$\pm$0.03  \\
 2973894 & $-0.59\pm$0.02 & +0.02$\pm$0.03 & +0.28$\pm$0.02 & +0.32$\pm$0.00 & +0.27$\pm$0.02 & +0.16$\pm$0.02 & +0.20$\pm$0.00 & +0.23$\pm$0.04  \\
 4446181 & $-0.26\pm$0.02 & +0.08$\pm$0.01 & +0.19$\pm$0.02 & +0.37$\pm$0.00 & +0.24$\pm$0.03 & +0.14$\pm$0.02 & +0.23$\pm$0.01  & +0.20$\pm$0.04 \\
 4751953 & $-0.25\pm$0.02 & +0.03$\pm$0.02 & +0.08$\pm$0.02 & +0.28$\pm$0.02 & +0.26$\pm$0.02 & +0.20$\pm$0.01 & +0.23$\pm$0.01 & +0.19$\pm$0.03 \\
 4661299 & $-0.41\pm$0.02 &  +0.14$\pm$0.02 & +0.16$\pm$0.02 & +0.53$\pm$0.00 & +0.29$\pm$0.03 & +0.21$\pm$0.01 & +0.32$\pm$0.02  & +0.24$\pm$0.04  \\
 4548530 & $-0.18\pm$0.03 & $-0.09\pm$0.02 & +0.11$\pm$0.01 & +0.24$\pm$0.02 & +0.25$\pm$0.02 & +0.15$\pm$0.03 & +0.24$\pm$0.02  & +0.19$\pm$0.04\\
10095427 & $-0.27\pm$0.02 & +0.06$\pm$0.03 & +0.18$\pm$0.01 & +0.31$\pm$0.02 & +0.27$\pm$0.02 & +0.22$\pm$0.02 & +0.31$\pm$0.01 & +0.25$\pm$0.03  \\
 4480358 & $-0.58\pm$0.03 & $-0.03\pm$0.02 & +0.33$\pm$0.01 & +0.36$\pm$0.01 & +0.27$\pm$0.03 & +0.09$\pm$0.02 & +0.20$\pm$0.03  & +0.22$\pm$0.04\\
 3936823 & $-0.54\pm$0.03 & +0.01$\pm$0.02 & +0.32$\pm$0.02 & +0.27$\pm$0.01 & +0.30$\pm$0.02 & +0.21$\pm$0.03 & +0.32$\pm$0.01  & +0.29$\pm$0.04 \\
 10586902 & $-0.31\pm$0.01  &  $-0.01\pm$0.00 &  +0.21$\pm$0.01 &  +0.22$\pm$0.00 &  +0.19$\pm$0.02 &  +0.08$\pm$0.01 &  +0.11$\pm$0.01 & +0.15$\pm$0.03\\
   9157260 &  $-0.21\pm$0.02 &   +0.05$\pm$0.01 &  +0.15$\pm$0.01 &  +0.15$\pm$0.00 &  +0.15$\pm$0.02 &  +0.08$\pm$0.01 &  +0.22$\pm$0.00 & +0.15$\pm$0.03\\
 4844527 &  $-0.29\pm$0.01 &   +0.04$\pm$0.02 &  +0.24$\pm$0.01 &  +0.33$\pm$0.00 &  +0.29$\pm$0.02 &  +0.12$\pm$0.02 &     $\ldots$ & +0.22$\pm$0.03 \\
  10463137 &  $-0.32\pm$0.02 &   $-0.05\pm$0.01 &  +0.19$\pm$0.01 &  +0.13$\pm$0.00 &  +0.17$\pm$0.03 &  +0.12$\pm$0.02 &     $\ldots$  &  +0.16$\pm$0.03 \\
  11870991 &  $-0.30\pm$0.01 &   +0.01$\pm$0.01 &  +0.22$\pm$0.01 &  +0.18$\pm$0.01 &  +0.20$\pm$0.02 &  +0.01$\pm$0.00 &  +0.02$\pm$0.01 & +0.11$\pm$0.03 \\

  \multicolumn{9}{c}{$\alpha$-normal} \\
4350501 &  $-0.04\pm$0.02  &  +0.01$\pm$0.01  &  +0.13$\pm$0.01 &  +0.19$\pm$0.02 &  +0.20$\pm$0.02 &  $-0.02\pm$0.02 &  +0.02$\pm$0.02 & +0.08$\pm$0.04 \\
9269081 &  $-0.04\pm$0.02  &  $-0.02\pm$0.01  &  +0.08$\pm$0.02 &  +0.22$\pm$0.02 &  +0.19$\pm$0.02 &  +0.06$\pm$0.02 &  +0.05$\pm$0.00  & +0.09$\pm$0.04\\
11445818 &  $-0.04\pm$0.01  &   $-0.04\pm$0.01 &  $-0.03\pm$0.01 &  +0.01$\pm$0.00 &  +0.05$\pm$0.01 &  $-0.07\pm$0.01 &  +0.10$\pm$0.00 & +0.01$\pm$0.03\\
3833399 &  +0.06$\pm$0.01  &   $-0.06\pm$0.01 & $-0.05\pm$0.00 &  $-0.09\pm$0.00 &  +0.05$\pm$0.01 &  $-0.08\pm$0.00 &  $-0.01\pm$0.01 & $-$0.02$\pm$0.03 \\
9761625 &  $-0.05\pm$0.01  &   +0.00$\pm$0.02  &  +0.03$\pm$0.01 &  +0.24$\pm$0.00 &  +0.16$\pm$0.01 &  +0.03$\pm$0.01 &  +0.06$\pm$0.01 & +0.07$\pm$0.03 \\
3455760 &  +0.00$\pm$0.01  &   $-0.02\pm$0.01  & +0.01$\pm$0.01  & $-0.08\pm$0.00 &  +0.01$\pm$0.01 &  $-0.04\pm$0.01 &  $-0.08\pm$0.00 & $-$0.02$\pm$0.03\\ 
9002884 &  $-0.15\pm$0.01  &   $-0.03\pm$0.01  & $-0.01\pm$0.01  &  +0.09$\pm$0.00 &  +0.09$\pm$0.01 &  +0.10$\pm$0.01 &  +0.10$\pm$0.00 & +0.07$\pm$0.03\\
2142095 & $-0.27\pm$0.01  &   +0.05$\pm$0.01 &  +0.12$\pm$0.01 &  $-0.09\pm$0.00 &  +0.18$\pm$0.01 &  +0.07$\pm$0.01 &  +0.01$\pm$0.01 & +0.09$\pm$0.03 \\
8611114 &  $-0.11\pm$0.02 &   $-0.10\pm$0.02 &  +0.07$\pm$0.00 &  $-0.05\pm$0.03 &  +0.10$\pm$0.02 &  +0.09$\pm$0.01 &  $-0.05\pm$0.01 & +0.05$\pm$0.03 \\
7594865 &  $-0.13\pm$0.02 &   $-0.08\pm$0.02 &  +0.11$\pm$0.01 &  +0.07$\pm$0.02 &  +0.19$\pm$0.02 &  +0.05$\pm$0.00 &  +0.00$\pm$0.01 & +0.09$\pm$0.03 \\
1432587  &  $-0.11\pm$0.02  &  $-0.08\pm$0.01 &  +0.13$\pm$0.00 &  +0.15$\pm$0.01 &  +0.13$\pm$0.01 &  +0.03$\pm$0.01 &  +0.04$\pm$0.02  & +0.08$\pm$0.03\\
3658136 &  $-0.01\pm$0.01  &  $-0.06\pm$0.01 &  +0.02$\pm$0.00 &  +0.03$\pm$0.00 &  +0.05$\pm$0.01 &  +0.01$\pm$0.02 &  +0.09$\pm$0.02  & +0.04$\pm$0.03\\
10880958 &  +0.03$\pm$0.02  & $-0.02\pm$0.02 &  $-0.05\pm$0.00 &  +0.12$\pm$0.04 &  +0.01$\pm$0.02 &  $-0.03\pm$0.01 &  $-0.18\pm$0.00 & $-$0.06$\pm$0.03\\
9143924 &  $-0.07\pm$0.01  &   $-0.01\pm$0.02 &  +0.04$\pm$0.00 &  +0.06$\pm$0.00 &  +0.10$\pm$0.02 &  $-0.04\pm$0.00 &  +0.01$\pm$0.01 & +0.03$\pm$0.03 \\
9605294 &  +0.04$\pm$0.01  &   +0.02$\pm$0.01 &  +0.05$\pm$0.00 &  $-0.10\pm$0.00 &  +0.04$\pm$0.02 &  $-0.02\pm$0.02 &  $-0.11\pm$0.02  & $-$0.01$\pm$0.03\\
\hline
\end{tabular}
\end{table*}

\begin{table*}
\centering
\caption{CNO abundances in $\alpha$-rich young and old red giants along with reference solar abundances. The combined random and systematic uncertainties in our measurements of [C/H], [N/H] and [O/H] are 0.06, 0.05 and 0.09 dex respectively. }
\label{cno_abundances}
\begin{tabular}{lcccccc}   \hline
   Sun     & 8.37  &  7.98 & 8.73  \\ \hline
\multicolumn{1}{c}{KIC} &\multicolumn{1}{c}{[C/H]} &\multicolumn{1}{c}{[N/H] } & \multicolumn{1}{c}{[O/H]}  & \multicolumn{1}{c}{N/C}  & \multicolumn{1}{c}{N/O} & \multicolumn{1}{c}{[C/N]}  \\ \hline
  \multicolumn{6}{c}{$\alpha$-rich (young)} \\
   4169517 & $-0.30$ & $-0.23$ & $-0.20$ & 0.5$\pm$0.2 & 0.17$\pm$0.09 & $-0.07$\\
   3946701 & $-0.35$ &  +0.29 &  +0.25 & 1.8$\pm$0.6 & 0.19$\pm$0.10 & $-0.64$\\
   8172784 & $-0.35$ &  +0.18 &  +0.20 & 1.4$\pm$0.5 & 0.17$\pm$0.09 & $-0.53$\\
   4136835 & $-0.40$ &  $\,\,\,\,$0.00 & +0.10 & 1.0$\pm$0.3 & 0.14$\pm$0.07 & $-0.40$\\
   4143460 & $-0.15$ & $-0.16$ &  +0.12 & 0.4$\pm$0.1 & 0.09$\pm$0.05 & $+0.01$\\
   6664950 & $-0.40$ &  +0.07 & $-0.05$ & 1.2$\pm$0.4 & 0.23$\pm$0.12 & $-0.47$\\
   8539201 & $-0.40$ & $\,\,\,\,0.00$ & $-0.10$ & 1.0$\pm$0.3 & 0.22$\pm$0.12 & $-0.40$\\
   5795626 & $-0.45$ & $-0.32$ & $-0.05$ & 0.5$\pm$0.2 & 0.10$\pm$0.05 & $-0.13$\\
  10096113 & $-0.34$ & $-0.20$ &  +0.10 & 0.6$\pm$0.2 & 0.09$\pm$0.05 & $-0.14$\\
   3973813 & $-0.45$ & $-0.08$ & $-0.05$ & 1.0$\pm$0.3 & 0.17$\pm$0.09 & $-0.37$\\
   1163359 & $-0.45$ &  +0.13 &  $\,\,\,\,$0.00 & 1.5$\pm$0.5 & 0.24$\pm$0.13 & $-0.58$\\
  12300740 & $-0.60$ &  +0.10 &  +0.20 & 2.0$\pm$0.7 & 0.16$\pm$0.07 & $-0.70$\\
   5956977 & $-0.40$ &  $\,\,\,\,$0.00 & $\,\,\,\,$ 0.00 & 1.0$\pm$0.3 & 0.18$\pm$0.09 & $-0.40$ \\
  11717920 & $-0.34$ & $-0.22$ & $-0.05$ & 0.5$\pm$0.2 & 0.12$\pm$0.06 & $-0.12$\\
   6837256 & $-0.25$ & $-0.27$ &  +0.12 & 0.4$\pm$0.1 & 0.07$\pm$0.04 & $+0.02$\\
   9821622 & $-0.20$ & $-0.20$ &  +0.05 & 0.4$\pm$0.1 & 0.10$\pm$0.05 & $\,\,\,\,0.00$\\
   4143460 & $-0.15$ & $-0.16$ &  +0.12  & 0.4$\pm$0.1 & 0.09$\pm$0.05 & $+0.01$\\
  11394905 & $-0.35$ & $-0.01$ & $-0.07$ & 0.9$\pm$0.3 & 0.20$\pm$0.11 & $-0.34$\\
  11823838 & $-0.30$ & $-0.03$ &  +0.05 & 0.8$\pm$0.3 & 0.15$\pm$0.08 & $-0.27$\\
   5512910 & $-0.32$ & $-0.17$ & $-0.05$ & 0.6$\pm$0.2 & 0.13$\pm$0.07 & $-0.15$\\
  10525475 & $-0.20$ &  +0.02 & $-0.06$ & 0.7$\pm$0.2 & 0.21$\pm$0.11 & $-0.22$\\

 \multicolumn{6}{c}{$\alpha$-rich (old)} \\
   3849996 & $-0.25$ & $-0.42$ & $-0.20$ & 0.3$\pm$0.1 & 0.11$\pm$0.06 & $+0.17$ \\
   7337994 & $-0.22$ & $-0.37$ & $-0.20$ & 0.3$\pm$0.1 & 0.12$\pm$0.06 & $+0.25$\\
   2973894 &  $\,\,\,\,$0.00 & $-0.02$ &  +0.05 & 0.4$\pm$0.1 & 0.15$\pm$0.08 & $+0.02$\\
   4446181 & $-0.40$ & $-0.31$ & $-0.20$ & 0.5$\pm$0.2 & 0.14$\pm$0.07 & $-0.09$\\
   4751953 & $-0.32$ & $-0.32$ & $-0.30$ & 0.4$\pm$0.1 & 0.17$\pm$0.09 & $\,\,\,\,0.00$\\
   4661299 & $-0.20$ & $-0.32$ & $-0.20$ & 0.3$\pm$0.1 & 0.13$\pm$0.07 & $+0.12$\\
   4548530 & $-0.32$ & $-0.34$ & $-0.20$ & 0.4$\pm$0.1 & 0.13$\pm$0.07 & +0.02\\
  10095427 & $-0.30$ & $-0.02$ &  +0.05 & 0.8$\pm$0.3 & 0.15$\pm$0.08 & $-0.28$\\
   4480358 & $-0.30$ & $-0.37$ &  +0.35 & 0.3$\pm$0.1 & 0.04$\pm$0.02 & $+0.07$\\
   3936823 & $-0.40$ & $-0.31$ & $-0.10$ & 0.5$\pm$0.2 & 0.11$\pm$0.06 & $-0.09$\\ 
  10586902 & $-0.30$ & $-0.18$ & $-0.25$ & 0.5$\pm$0.2 & 0.21$\pm$0.11 & $-0.12$ \\
   9157260 & $-0.15$ & $-0.21$ & $-0.15$ & 0.4$\pm$0.1 & 0.15$\pm$0.08 & $+0.06$\\
 4844527 & $-0.20$ & $-0.23$ & $-0.05$ & 0.4$\pm$0.1 & 0.12$\pm$0.06 & $ +0.03$\\
  10463137 & $-0.45$ & $-0.13$ & $-0.10$ & 0.9$\pm$0.3 & 0.17$\pm$0.09 & $-0.32$\\
  11870991 & $-0.40$ & $-0.12$ & $-0.25$ & 0.8$\pm$0.3 & 0.24$\pm$0.13 & $-0.28$\\

  \multicolumn{6}{c}{$\alpha$-normal} \\
   4350501 & $-0.15$ &  +0.04 & $-0.05$ & 0.6$\pm$0.2 & 0.22$\pm$0.12 & $-0.19$\\
    9269081 & $-0.20$ &  +0.23 &  +0.12 & 1.1$\pm$0.4 & 0.23$\pm$0.12 & $-0.43$\\
    9002884 & $-0.22$ &  +0.06 &  +0.12 & 0.8$\pm$0.3 & 0.15$\pm$0.08 & $-0.28$\\
   9761625 & $-0.20$ &  +0.04 &  +0.12 & 0.7$\pm$0.2 & 0.15$\pm$0.08 & $-0.24$\\
  11445818 &  +0.02 &  +0.15 &  +0.10 & 0.5$\pm$0.2 & 0.20$\pm$0.11 & $-0.13$\\
   3455760 &  +0.05 &  +0.11 &  +0.06 & 0.5$\pm$0.2 & 0.20$\pm$0.11 & $-0.06$\\
   3833399 &  +0.10 &  +0.09 &  +0.07 & 0.4$\pm$0.1 & 0.19$\pm$0.10 & $+0.01$\\
   2142095 & $-0.20$ & $-0.15$ & $-0.25$ & 0.5$\pm$0.2 & 0.22$\pm$0.12 & $-0.05$\\
   8611114 & $-0.30$ & $-0.32$ & $-0.20$ & 0.4$\pm$0.1 & 0.13$\pm$0.07 & $+0.02$\\
   7594865 & $-0.35$ & $-0.13$ & $-0.10$ & 0.7$\pm$0.2 & 0.17$\pm$0.09 & $-0.22$\\
  1432587 & $-0.20$ & $-0.08$ &  +0.05 & 0.5$\pm$0.2 & 0.13$\pm$0.07 & $-0.28$\\
   3658136 & $-0.25$ &  +0.11 &  +0.03 & 0.9$\pm$0.3 & 0.21$\pm$0.11 & $-0.36$\\
  10880958 & $-0.15$ &  +0.02 & $-0.11$ & 0.6$\pm$0.2 & 0.24$\pm$0.13 & $-0.17$\\
   9143924 & $-0.18$ &  +0.06 &  +0.05 & 0.7$\pm$0.2 & 0.18$\pm$0.10 & $-0.24$\\
   9605294 & $-0.10$ &  +0.02 & $-0.10$ & 0.5$\pm$0.2 & 0.23$\pm$0.12 & $-0.12$\\   
   \hline

\end{tabular}
\end{table*}

\begin{table*}
\centering
\caption{Sensitivity of derived abundances to uncertainties in stellar parameters for the star KIC\,6664950 with $T_{\rm eff}\
=$4800 K, log~$g=$2.4 dex and $\xi_{t}=$ 1.7 km-s$^{-1}$. Each number refers to the difference in abundances obtained while v\
arying one stellar parameter both increasing and decreasing them (results of both are shown separated by `/'), while keeping \
the other parameters unchanged.}
\label{abu_sensitivity}
\begin{tabular}{cccccc}   \hline
\multicolumn{1}{c}{Species} &\multicolumn{1}{c}{$T_{\rm eff}\pm$50~K} &\multicolumn{1}{c}{$\log\,g\pm$0.2} & \multicolumn{1}{c}{$\xi_{t}\pm$0.2~km\,s$^{-1}$} &
\multicolumn{1}{c}{[M/H]$\pm$0.1} & \multicolumn{1}{l}{ } \\ \cline{2-5}

\multicolumn{1}{c}{(X)}&\multicolumn{1}{c}{$\delta_{X}$} &\multicolumn{1}{c}{$\delta_{X}$} & \multicolumn{1}{c}{$\delta_{X}$}\
 &
\multicolumn{1}{c}{$\delta_{X}$}  &\multicolumn{1}{c}{$\sigma_{2}$}  \\ \hline


 $[$C/Fe$]$  & $+0.04/-0.01$  &$+0.03/+0.02$   &$+0.02/+0.01$  &$+0.05/-0.03$   & 0.05  \\
 $[$N/Fe$]$  & $-0.02/-0.01$  &$+0.03/-0.01$   &$+0.08/+0.02$  &$+0.04/$ $0.00$  & 0.06  \\
 $[$O/Fe$]$  & $+0.01/-0.01$  &$-0.02/-0.04$   &$-0.01/-0.02$  &$+0.09/-0.07$   & 0.09  \\
$[$Mg/Fe$]$  & $+0.01/$ $0.00$ &$+0.01/+0.01$   &$+0.01/-0.02$  &$-0.01/$ $0.00$  & 0.02  \\
$[$Al/Fe$]$  & $+0.02/-0.02$  &$-0.02/+0.04$   &$-0.02/+0.01$  & $0.00/\,0.00$ & 0.04  \\
$[$Si/Fe$]$  & $-0.01/+0.02$  &$+0.03/\,0.00$  &$-0.01/+0.01$  &$+0.02/-0.01$   & 0.03  \\
$[$Ca/Fe$]$  & $-0.01/-0.01$  &$+0.01/+0.01$   &$+0.02/-0.02$  &$-0.01/\,0.00$  & 0.02  \\
$[$Ti/Fe$]$  & $+0.05/-0.04$  &$+0.02/\,0.00$  &$+0.02/-0.03$  &$-0.01/+0.01$   & 0.05  \\
$[$Fe/H$]$   & $+0.02/-0.03$  &$-0.03/\,0.00$  &$-0.04/+0.04$  &$+0.01/-0.02$   & 0.05  \\
$[$Ni/Fe$]$  & $\,0.00/+0.02$ &$+0.05/-0.01$   &$+0.02/-0.01$  & $0.00/-0.01$   & 0.03  \\

\hline
\end{tabular}
\end{table*}

\begin{table*}
\centering
\caption{Equivalent width measurements for individual lines in m\AA\ for KIC4169517, KIC4169517, KIC8172784, KIC4136835, KIC4143460, KIC6664950, KIC8539201 and KIC5795626.}
\label{EWs0}
\begin{tabular}{lccccccccc}  
\hline
\multicolumn{1}{c}{Element} &\multicolumn{1}{c}{$\lambda$ [\AA\ ]} &\multicolumn{1}{c}{KIC4169517} & \multicolumn{1}{c}{KIC4169517} & \multicolumn{1}{c}{KIC8172784}& \multicolumn{1}{c}{KIC4136835} &  \multicolumn{1}{c}{KIC4143460}& \multicolumn{1}{c}{KIC6664950}& \multicolumn{1}{c}{KIC8539201}& \multicolumn{1}{c}{KIC5795626}\\
\hline
Fe & 15176.72 &  &  &  &  &  &  &  & 23.22\\
Fe & 15301.56 & 43.88 &  & 45.35 & 32.55 & 62.50 & 54.77 & 29.05 & 29.85\\
Fe & 15394.67 &  &  &  &  &  &  &  & 104.50\\
Fe & 15395.72 &  &  &  & 86.73 &  &  & 102.10 & 87.54\\
Fe & 15485.45 &  &  &  &  &  &  &  & \\
Fe & 15524.30 & 14.45 & 16.72 &  &  &  &  &  & \\
Fe & 15534.26 & 113.30 & 110.40 & 120.10 & 86.90 & 129.40 & 127.50 & 93.60 & \\
Fe & 15648.52 & 95.78 & 105.50 & 101.90 & 80.35 &  &  & 81.90 & \\
Fe & 15652.87 & 75.10 &  &  &  &  &  &  & \\
Fe & 15733.51 &  &  &  &  &  &  &  & \\
Fe & 15761.31 & 67.71 &  &  &  &  &  &  & \\
Fe & 15774.07 & 124.50 & 121.20 & 129.40 & 93.92 & 145.40 & 140.50 &  & \\
Fe & 15609.04 &  &  &  &  &  &  &  & \\
Fe & 15904.35 & 109.70 &  & 132.70 & 97.50 &  & 139.30 & 101.70 & 89.70\\
Fe & 15906.04 & 139.80 &  & 145.40 & 105.50 & 161.80 & 159.20 & 122.90 & 110.70\\
Fe & 15934.02 & 48.76 & 45.93 & 53.67 & 33.90 & 65.10 & 54.40 & 35.49 & 32.58\\
Fe & 15952.63 & 21.22 & 20.80 & 26.54 & 15.87 & 35.33 &  &  & 13.36\\
Fe & 15964.87 & 108.10 &  &  &  &  &  &  & \\
Fe & 15980.73 & 170.30 & 178.60 & 181.90 & 140.50 &  &  & 159.90 & 140.70\\
Fe & 15997.74 & 62.24 &  &  &  &  &  &  & \\
Fe & 16040.65 & 136.10 & 140.20 &  &  &  &  & 120.10 & 110.50\\
Fe & 16102.41 & 152.60 & 155.70 & 170.50 & 132.60 & 183.40 & 183.80 & 150.10 & 121.80\\
Fe & 16180.93 & 105.70 &  &  & 85.40 & 131.50 & 119.90 & 83.60 & 85.24\\
Fe & 16198.51 & 130.00 & 137.50 & 139.30 & 110.20 & 159.10 & 151.50 & 124.00 & 100.50\\
Fe & 16225.64 & 81.74 &  &  & 68.20 & 107.30 & 105.90 & 71.50 & 65.10\\
Fe & 16292.85 & 67.90 & 79.52 &  & 54.30 &  &  & 58.97 & 48.80\\
Fe & 16324.46 & 127.70 & 136.40 & 137.70 & 100.80 & 159.40 & 149.80 & 119.50 & 100.80\\
Fe & 16331.53 &  &  &  &  &  &  &  & \\
Ni & 16310.51 & 84.40 & 81.65 & 81.60 & 77.60 & 103.90 & 103.80 & 77.70 & 60.70\\
Ni & 16363.11 & 132.90 & 129.40 & 131.70 & 117.30 &  & 161.90 & 132.90 & 112.70\\
Ni & 16388.75 & 19.50 &  &  &  &  &  &  & \\
Ni & 16584.44 & 41.20 & 36.90 &  &  & 59.50 & 52.20 &  & 29.50\\
Ni & 16673.71 & 38.92 &  &  & 38.30 & 57.60 & 57.79 & 33.70 & 31.35\\
Ni & 16815.47 & 35.80 & 34.84 & 33.70 & 32.33 & 54.80 & 46.60 & 30.00 & 28.35\\
Mg & 15879.56 & 155.30 & 162.70 & 155.50 & 141.70 & 175.50 & 174.20 & 136.10 & 125.30\\
Mg & 15886.19 & 115.70 & 113.90 & 106.50 & 99.91 & 134.50 & 124.30 & 96.50 & 84.04\\
Al & 16718.98 & 298.30 & 294.40 & 332.50 & 275.60 & 342.50 & 337.60 & 292.80 & 273.20\\
Al & 16750.57 &  &  &  &  &  &  &  & \\
Al & 16763.35 & 189.30 & 181.60 & 204.50 & 159.80 & 225.20 & 220.90 & 168.70 & 160.60\\
Si & 15557.79 & 187.20 & 175.10 & 180.30 & 159.20 & 200.70 & 205.20 & 171.20 & 166.30\\
Si & 16060.02 & 242.30 & 226.30 & 244.70 & 217.20 & 255.50 & 258.50 & 226.70 & 210.10\\
Si & 16163.71 & 183.00 & 171.20 & 171.40 & 154.50 & 196.90 & 192.30 & 160.20 & 165.10\\
Si & 16215.71 & 222.10 & 211.90 & 209.10 & 189.50 & 240.70 & 239.90 & 208.30 & 196.90\\
Si & 16241.87 & 191.70 & 179.40 & 186.50 & 170.90 & 208.70 & 208.60 & 178.50 & 172.60\\
Si & 16828.18 & 151.30 & 139.90 & 147.40 & 125.60 & 168.50 & 171.40 & 139.70 & 136.30\\
Ca & 16150.76 & 83.30 & 82.70 & 90.49 & 84.90 & 102.60 & 96.90 & 70.50 & 61.55\\
Ca & 16155.27 & 41.90 & 42.78 & 44.40 & 39.89 & 49.65 & 51.80 & 29.01 & 24.60\\
Ca & 16197.04 & 134.50 & 131.50 & 141.50 & 121.00 & 148.20 & 151.70 & 114.70 & 99.60\\
Ti & 15602.84 &  & 24.55 &  & 13.60 & 32.45 & 28.50 & 14.45 & 11.90\\
Ti & 15715.57 & 49.20 & 86.60 & 68.50 & 54.60 &  &  & 61.77 & 47.76\\
Ti & 15543.78 & 59.65 & 95.12 & 73.45 & 61.40 & 110.10 & 102.18 & 67.00 & \\
\hline
\end{tabular}
\end{table*}

\begin{table*}
\centering
\caption{Equivalent width measurements for individual lines in m\AA\ for KIC10096113, KIC3973813, KIC1163359, KIC12300740, KIC5956977, KIC11717920, KIC6837256, KIC9821622.}
\label{EWs1}
\begin{tabular}{lccccccccc}  
\hline
\multicolumn{1}{c}{Element} &\multicolumn{1}{c}{$\lambda$ [\AA\ ]} &\multicolumn{1}{c}{KIC10096113} & \multicolumn{1}{c}{KIC3973813} & \multicolumn{1}{c}{KIC1163359}& \multicolumn{1}{c}{KIC12300740} &  \multicolumn{1}{c}{KIC5956977}& \multicolumn{1}{c}{KIC11717920}& \multicolumn{1}{c}{KIC6837256}& \multicolumn{1}{c}{KIC9821622}\\
\hline
Fe & 15176.72 &  &  &  &  &  &  &  & \\
Fe & 15301.56 &  & 52.50 & 55.55 & 29.71 & 35.10 &  & 33.13 & \\
Fe & 15394.67 &  &  & 155.50 & 109.50 & 131.70 & 114.80 &  & \\
Fe & 15395.72 & 96.80 &  & 141.20 & 98.75 & 123.20 &  &  & \\
Fe & 15485.45 &  & 25.33 &  &  &  &  &  & \\
Fe & 15524.30 &  &  &  &  &  &  &  & \\
Fe & 15534.26 & 87.10 &  & 138.60 &  & 105.40 &  & 117.50 & \\
Fe & 15648.52 &  & 114.60 & 131.90 &  & 107.60 &  &  & \\
Fe & 15652.87 &  & 85.50 &  &  &  &  &  & \\
Fe & 15733.51 & 20.97 & 42.31 &  &  &  &  &  & \\
Fe & 15761.31 &  &  &  &  &  &  &  & \\
Fe & 15774.07 & 95.70 &  &  & 95.50 & 119.50 & 105.60 & 117.80 & \\
Fe & 15609.04 &  &  &  &  &  &  &  & \\
Fe & 15904.35 & 91.72 &  &  &  &  & 98.50 & 114.75 & 128.50\\
Fe & 15906.04 & 108.50 &  &  &  &  & 124.20 & 145.50 & 151.70\\
Fe & 15934.02 &  & 65.36 & 62.42 & 34.90 & 48.10 & 39.07 & 41.90 & 50.90\\
Fe & 15952.63 &  &  & 33.54 &  &  &  & 18.95 & \\
Fe & 15964.87 &  & 133.20 &  &  &  &  &  & \\
Fe & 15980.73 &  & 191.80 & 195.00 & 144.10 & 181.90 & 158.70 & 183.60 & \\
Fe & 15997.74 &  & 82.60 &  &  &  &  &  & \\
Fe & 16040.65 &  &  & 165.50 & 119.90 & 148.30 & 118.10 & 137.90 & \\
Fe & 16102.41 & 134.20 & 182.60 & 181.30 & 137.00 & 159.10 & 148.40 & 175.50 & 178.35\\
Fe & 16180.93 & 82.45 & 119.50 & 125.70 &  & 101.50 &  & 99.02 & 113.20\\
Fe & 16198.51 & 102.70 &  & 172.60 & 110.50 & 140.20 & 121.00 & 130.80 & 151.70\\
Fe & 16225.64 &  & 105.50 & 105.50 & 62.75 & 85.40 & 63.64 & 80.75 & 96.50\\
Fe & 16292.85 &  & 92.94 & 102.50 & 51.40 &  & 61.10 &  & 82.79\\
Fe & 16324.46 &  & 156.60 & 162.40 & 107.90 & 137.40 & 117.40 & 132.00 & 145.70\\
Fe & 16331.53 &  & 81.32 &  &  &  &  &  & \\
Ni & 16310.51 &  & 103.50 & 103.20 & 70.45 & 107.10 & 80.80 & 90.50 & \\
Ni & 16363.11 & 127.30 & 158.90 & 159.40 & 117.50 & 168.10 & 139.60 & 159.70 & 158.50\\
Ni & 16388.75 &  &  &  & 15.64 & 28.40 &  &  & \\
Ni & 16584.44 & 39.40 &  & 58.40 &  & 50.90 &  & 41.68 & 55.60\\
Ni & 16673.71 &  & 53.25 & 50.45 & 30.60 & 50.91 & 41.46 & 43.79 & 51.00\\
Ni & 16815.47 &  & 52.31 &  &  &  &  & 35.35 & \\
Mg & 15879.56 & 145.20 & 168.90 & 169.50 & 142.80 & 170.30 & 138.90 & 150.50 & 200.20\\
Mg & 15886.19 & 111.30 & 131.60 & 125.60 & 102.60 & 128.80 & 101.40 & 107.80 & 153.30\\
Al & 16718.98 & 299.00 & 363.10 & 339.60 & 285.80 & 344.90 & 311.70 & 327.80 & 350.50\\
Al & 16750.57 & 373.50 &  &  &  &  &  &  & \\
Al & 16763.35 &  & 241.60 & 225.00 & 164.90 & 217.90 & 189.90 & 194.90 & 222.40\\
Si & 15557.79 & 164.40 &  & 182.50 & 162.20 & 182.40 & 167.50 & 172.70 & \\
Si & 16060.02 & 209.20 & 236.80 & 243.80 & 212.50 & 241.10 & 221.70 & 225.00 & 255.50\\
Si & 16163.71 & 160.60 & 178.50 & 185.10 & 159.60 & 179.00 & 166.60 & 168.90 & 182.60\\
Si & 16215.71 & 193.20 & 216.70 &  & 194.70 &  & 206.50 & 215.55 & 229.50\\
Si & 16241.87 &  &  &  & 171.50 & 185.90 & 186.90 & 190.20 & 195.80\\
Si & 16828.18 & 131.20 & 155.50 & 164.10 & 135.10 & 152.70 & 145.80 & 147.40 & 159.80\\
Ca & 16150.76 & 75.60 & 124.60 & 111.30 & 77.00 & 92.23 & 82.60 & 88.34 & 115.30\\
Ca & 16155.27 & 32.70 & 61.81 & 55.55 &  & 42.13 &  & 37.81 & 58.90\\
Ca & 16197.04 & 108.90 & 170.50 & 161.70 & 109.50 & 135.40 & 120.70 & 141.90 & 162.50\\
Ti & 15602.84 &  &  & 41.00 & 18.50 & 31.50 & 23.54 & 24.55 & 34.50\\
Ti & 15715.57 & 56.70 & 141.20 & 127.30 & 78.30 & 110.50 & 96.30 & 99.20 & 110.40\\
Ti & 15543.78 &  & 146.80 & 135.50 & 85.50 & 111.50 & 92.91 & 104.50 & \\
\hline
\end{tabular}
\end{table*}

\begin{table*}
\centering
\caption{Equivalent width measurements for individual lines in m\AA\ for KIC4143460, KIC11394905, KIC11823838, KIC5512910, KIC10525475, KIC3849996, KIC7337994, KIC2973894.}
\label{EWs2}
\begin{tabular}{lccccccccc}  
\hline
\multicolumn{1}{c}{Element} &\multicolumn{1}{c}{$\lambda$ [\AA\ ]} &\multicolumn{1}{c}{KIC4143460} & \multicolumn{1}{c}{KIC11394905} & \multicolumn{1}{c}{KIC11823838}& \multicolumn{1}{c}{KIC5512910} &  \multicolumn{1}{c}{KIC10525475}& \multicolumn{1}{c}{KIC3849996}& \multicolumn{1}{c}{KIC7337994}& \multicolumn{1}{c}{KIC2973894}\\
\hline
Fe & 15176.72 &  &  &  &  &  &  &  & \\
Fe & 15301.56 & 62.50 &  &  &  &  & 47.45 & 58.50 & 23.72\\
Fe & 15394.67 &  &  &  &  &  &  &  & \\
Fe & 15395.72 &  &  &  &  &  &  &  & \\
Fe & 15485.45 &  &  &  &  &  &  &  & \\
Fe & 15524.30 &  &  &  &  &  &  &  & \\
Fe & 15534.26 & 129.40 & 109.50 & 126.90 & 115.90 & 142.50 &  &  & \\
Fe & 15648.52 &  &  &  &  &  &  &  & \\
Fe & 15652.87 &  &  &  &  &  &  &  & \\
Fe & 15733.51 &  &  &  &  &  &  &  & \\
Fe & 15761.31 &  &  &  &  &  &  &  & \\
Fe & 15774.07 & 145.40 & 118.50 & 128.90 & 124.40 & 145.90 &  &  & \\
Fe & 15609.04 &  &  &  &  &  &  &  & \\
Fe & 15904.35 &  &  &  &  &  & 121.90 &  & \\
Fe & 15906.04 & 161.80 & 137.30 & 149.60 & 139.50 &  & 135.60 &  & 112.80\\
Fe & 15934.02 & 65.10 & 48.15 & 55.11 & 49.70 & 67.80 &  & 60.90 & 28.80\\
Fe & 15952.63 & 35.33 &  & 26.74 &  & 33.90 & 23.40 & 31.50 & \\
Fe & 15964.87 &  &  &  &  &  &  &  & \\
Fe & 15980.73 &  &  &  &  &  & 170.60 &  & \\
Fe & 15997.74 &  &  &  &  &  &  &  & \\
Fe & 16040.65 &  &  &  &  &  & 142.60 & 145.80 & 118.90\\
Fe & 16102.41 & 183.40 & 157.50 &  & 156.10 & 191.60 &  & 167.70 & 138.90\\
Fe & 16180.93 & 131.50 & 109.30 & 116.30 & 109.20 & 140.20 & 108.90 & 117.70 & \\
Fe & 16198.51 & 159.10 & 135.00 & 154.90 & 136.70 & 172.30 & 135.50 & 142.50 & \\
Fe & 16225.64 & 107.30 & 88.45 & 98.30 &  & 115.70 & 89.17 & 101.50 & 59.55\\
Fe & 16292.85 &  &  &  &  &  & 76.77 & 85.70 & \\
Fe & 16324.46 & 159.40 & 133.80 & 148.00 & 135.30 & 163.50 & 133.50 & 148.60 & 107.50\\
Fe & 16331.53 &  &  &  &  &  &  &  & \\
Ni & 16310.51 & 103.90 & 87.70 & 88.50 & 80.52 & 113.50 & 86.80 & 101.45 & 63.02\\
Ni & 16363.11 &  &  &  &  &  & 138.20 &  & 123.40\\
Ni & 16388.75 &  &  &  &  &  &  &  & \\
Ni & 16584.44 & 59.50 & 48.88 & 43.60 &  & 62.50 &  & 55.21 & 29.91\\
Ni & 16673.71 & 57.60 & 48.51 & 41.43 & 42.72 & 61.90 & 41.25 & 58.50 & 29.10\\
Ni & 16815.47 & 54.80 & 44.83 &  & 36.39 & 56.50 & 37.45 & 49.44 & 22.11\\
Mg & 15879.56 & 175.50 & 169.30 & 166.10 & 141.30 & 187.60 & 155.60 & 164.30 & 121.40\\
Mg & 15886.19 & 134.50 & 128.50 & 121.90 & 102.20 & 138.30 & 110.98 & 123.50 & 74.75\\
Al & 16718.98 & 342.50 & 302.70 & 333.50 & 302.60 & 338.40 & 301.30 & 334.10 & 254.80\\
Al & 16750.57 &  &  &  &  &  &  &  & \\
Al & 16763.35 & 225.20 & 224.60 & 214.20 & 191.80 & 251.30 & 191.90 & 216.70 & 156.50\\
Si & 15557.79 & 200.70 & 182.50 & 185.20 &  &  & 186.90 & 202.70 & \\
Si & 16060.02 & 255.50 & 240.40 & 235.10 & 225.80 & 250.80 & 245.30 & 255.80 & 217.60\\
Si & 16163.71 & 196.90 & 174.50 & 180.45 & 169.30 & 190.70 & 185.90 & 194.40 & 165.90\\
Si & 16215.71 & 240.70 & 220.50 & 222.50 & 211.20 & 239.30 & 223.90 &  & \\
Si & 16241.87 & 208.70 & 193.80 & 192.70 & 185.80 & 201.80 & 200.80 & 205.30 & 178.70\\
Si & 16828.18 & 168.50 & 151.40 & 155.20 & 145.20 & 164.30 & 163.40 & 169.60 & 129.80\\
Ca & 16150.76 & 102.60 & 98.16 & 93.90 & 85.91 & 117.20 & 80.60 & 102.26 & 44.51\\
Ca & 16155.27 & 49.65 &  & 45.20 & 41.55 & 62.50 & 37.21 &  & 16.40\\
Ca & 16197.04 & 148.20 & 139.70 & 140.20 &  & 165.30 & 122.20 & 146.10 & \\
Ti & 15602.84 & 32.45 & 19.50 & 22.71 & 21.30 & 38.55 & 20.95 & 30.50 & 8.66\\
Ti & 15715.57 &  &  &  &  &  & 76.72 & 95.70 & 79.20\\
Ti & 15543.78 & 110.10 & 81.28 & 92.80 &  &  & 80.60 & 102.90 & \\
\hline
\end{tabular}
\end{table*}

\begin{table*}
\centering
\caption{Equivalent width measurements for individual lines in m\AA\ for KIC4446181, KIC4751953, KIC4661299, KIC4548530, KIC10095427, KIC4480358, KIC3936823, KIC10586902.}
\label{EWs3}
\begin{tabular}{lccccccccc}  
\hline
\multicolumn{1}{c}{Element} &\multicolumn{1}{c}{$\lambda$ [\AA\ ]} &\multicolumn{1}{c}{KIC4446181} & \multicolumn{1}{c}{KIC4751953} & \multicolumn{1}{c}{KIC4661299}& \multicolumn{1}{c}{KIC4548530} &  \multicolumn{1}{c}{KIC10095427}& \multicolumn{1}{c}{KIC4480358}& \multicolumn{1}{c}{KIC3936823}& \multicolumn{1}{c}{KIC10586902}\\
\hline
Fe & 15176.72 &  &  &  &  &  &  &  & \\
Fe & 15301.56 & 44.94 & 52.20 & 34.45 & 58.69 & 48.47 & 33.49 & 36.60 & 47.28\\
Fe & 15394.67 &  &  &  &  &  &  &  & \\
Fe & 15395.72 &  &  &  &  &  &  &  & \\
Fe & 15485.45 &  &  &  &  &  &  &  & \\
Fe & 15524.30 &  &  &  &  &  &  &  & \\
Fe & 15534.26 &  &  &  &  &  &  &  & 118.90\\
Fe & 15648.52 &  &  &  &  &  &  &  & \\
Fe & 15652.87 &  &  &  &  &  &  &  & \\
Fe & 15733.51 &  &  &  &  &  &  &  & \\
Fe & 15761.31 &  &  &  &  &  &  &  & \\
Fe & 15774.07 &  &  &  &  &  &  &  & \\
Fe & 15609.04 &  &  &  &  &  &  &  & \\
Fe & 15904.35 & 125.70 & 128.50 &  &  &  & 103.85 & 107.40 & \\
Fe & 15906.04 & 148.80 & 148.10 &  &  &  & 132.50 & 140.70 & 151.80\\
Fe & 15934.02 & 53.20 & 54.86 & 38.80 & 62.05 & 55.19 & 35.90 & 39.14 & 54.95\\
Fe & 15952.63 &  & 28.02 & 19.55 & 34.47 & 25.58 & 17.45 &  & 24.70\\
Fe & 15964.87 &  &  &  &  &  &  &  & \\
Fe & 15980.73 &  &  &  &  &  &  &  & \\
Fe & 15997.74 &  &  &  &  &  &  &  & \\
Fe & 16040.65 & 142.10 & 145.50 & 127.80 &  & 151.80 & 127.45 & 146.80 & \\
Fe & 16102.41 & 164.30 & 162.10 & 145.30 & 188.60 & 170.50 & 154.80 & 156.40 & 173.50\\
Fe & 16180.93 & 118.60 &  & 94.03 & 129.40 & 119.70 & 88.46 & 100.50 & 115.70\\
Fe & 16198.51 & 146.10 & 149.10 & 121.50 & 160.80 & 149.30 & 128.90 & 137.60 & 149.60\\
Fe & 16225.64 & 90.40 & 92.79 &  & 110.10 & 100.40 & 73.45 & 80.95 & 92.60\\
Fe & 16292.85 & 80.81 & 82.10 &  & 92.35 & 86.45 & 58.82 & 69.70 & \\
Fe & 16324.46 & 142.90 & 145.40 & 121.40 & 159.80 & 148.20 & 133.90 & 143.50 & 148.30\\
Fe & 16331.53 &  &  &  &  &  &  &  & \\
Ni & 16310.51 & 102.10 & 98.90 & 91.40 & 98.85 & 105.80 & 72.30 &  & 92.20\\
Ni & 16363.11 & 159.90 & 152.50 & 143.60 & 151.80 & 160.50 & 129.55 & 141.30 & \\
Ni & 16388.75 &  &  &  &  &  &  &  & \\
Ni & 16584.44 &  &  &  & 51.91 & 54.40 & 32.45 & 43.30 & 48.80\\
Ni & 16673.71 & 55.40 &  & 49.58 & 51.60 & 54.30 & 30.78 & 36.30 & \\
Ni & 16815.47 & 51.45 & 51.22 & 40.89 & 49.35 & 48.80 & 29.95 & 33.60 & 42.10\\
Mg & 15879.56 & 169.70 & 158.70 & 134.00 & 177.60 & 181.90 & 131.50 & 180.90 & 155.90\\
Mg & 15886.19 & 122.70 & 113.50 & 97.94 & 131.30 & 133.10 & 87.84 & 128.10 & 110.10\\
Al & 16718.98 & 328.70 & 312.90 & 316.90 & 332.50 & 358.00 & 319.50 & 327.10 & 335.90\\
Al & 16750.57 &  &  &  &  &  &  &  & \\
Al & 16763.35 & 210.50 & 203.50 &  & 218.90 & 231.80 & 200.70 & 200.50 & 210.40\\
Si & 15557.79 & 186.50 & 187.40 & 188.70 &  &  &  &  & 170.30\\
Si & 16060.02 & 234.50 & 245.40 & 239.40 & 258.90 &  & 208.50 & 220.40 & 228.60\\
Si & 16163.71 & 184.75 & 185.50 & 178.80 & 196.70 & 176.00 & 156.20 & 172.35 & 163.50\\
Si & 16215.71 & 225.10 & 229.90 & 219.50 & 237.10 &  & 197.65 & 217.20 & 210.40\\
Si & 16241.87 & 198.30 & 196.50 & 200.70 & 207.40 & 195.90 & 172.45 & 190.50 & 182.60\\
Si & 16828.18 &  & 163.45 & 157.00 & 171.00 & 157.80 & 137.40 & 148.25 & 142.60\\
Ca & 16150.76 & 83.38 & 88.45 & 89.30 & 97.45 & 122.60 & 73.61 & 102.30 & 100.79\\
Ca & 16155.27 & 36.55 & 40.65 & 44.65 & 42.50 & 65.65 & 33.53 & 44.20 & 51.30\\
Ca & 16197.04 &  & 127.25 & 134.80 & 137.80 & 177.50 & 122.50 & 150.90 & 149.60\\
Ti & 15602.84 & 29.15 & 37.50 & 19.50 & 40.55 & 51.40 & 28.76 & 49.90 & 30.60\\
Ti & 15715.57 & 98.45 & 108.50 & 73.90 & 124.10 & 136.80 & 106.50 & 154.60 & \\
Ti & 15543.78 &  & 119.10 &  & 133.45 & 142.50 & 110.20 &  & 114.30\\
\hline
\end{tabular}
\end{table*}
       
\begin{table*}
\centering
\caption{Equivalent width measurements for individual lines in m\AA\ for KIC9157260, KIC4844527, KIC10463137, KIC11870991, KIC4350501, KIC9269081, KIC11445818, KIC3833399.}
\label{EWs4}
\begin{tabular}{lccccccccc}  
\hline
\multicolumn{1}{c}{Element} &\multicolumn{1}{c}{$\lambda$ [\AA\ ]} &\multicolumn{1}{c}{KIC9157260} & \multicolumn{1}{c}{KIC4844527} & \multicolumn{1}{c}{KIC10463137}& \multicolumn{1}{c}{KIC11870991} &  \multicolumn{1}{c}{KIC4350501}& \multicolumn{1}{c}{KIC9269081}& \multicolumn{1}{c}{KIC11445818}& \multicolumn{1}{c}{KIC3833399}\\
\hline
Fe & 15176.72 &  &  &  &  &  &  &  & \\
Fe & 15301.56 & 50.13 & 44.36 & 42.50 &  & 59.50 & 71.37 &  & \\
Fe & 15394.67 &  &  &  &  &  &  &  & \\
Fe & 15395.72 &  &  &  &  &  &  &  & \\
Fe & 15485.45 &  &  &  &  &  &  &  & \\
Fe & 15524.30 &  &  &  &  &  &  &  & \\
Fe & 15534.26 & 136.10 & 117.20 & 116.50 &  & 138.80 & 152.60 &  & \\
Fe & 15648.52 &  &  &  &  &  &  &  & \\
Fe & 15652.87 &  &  &  &  &  &  &  & \\
Fe & 15733.51 &  &  &  &  &  &  &  & \\
Fe & 15761.31 &  &  &  &  &  &  &  & \\
Fe & 15774.07 & 148.50 & 124.30 & 127.90 & 127.10 & 152.20 & 155.70 &  & 183.20\\
Fe & 15609.04 &  &  &  &  &  &  &  & \\
Fe & 15904.35 &  &  &  &  &  &  &  & \\
Fe & 15906.04 & 168.10 & 142.10 & 144.50 & 152.00 & 175.50 & 175.20 & 187.90 & 206.10\\
Fe & 15934.02 & 56.27 & 50.48 & 50.40 & 51.40 & 70.70 & 77.28 & 80.56 & 95.42\\
Fe & 15952.63 & 28.10 & 25.50 &  & 24.77 & 39.50 & 41.34 & 41.75 & 51.50\\
Fe & 15964.87 &  &  &  &  &  &  &  & \\
Fe & 15980.73 &  &  &  &  &  &  &  & \\
Fe & 15997.74 &  &  &  &  &  &  &  & \\
Fe & 16040.65 &  &  &  &  &  &  &  & \\
Fe & 16102.41 & 205.70 & 164.00 & 172.70 & 177.20 & 204.50 & 202.10 & 212.40 & 235.50\\
Fe & 16180.93 & 130.30 & 112.20 & 110.20 & 116.50 & 146.60 & 150.60 & 153.50 & 172.20\\
Fe & 16198.51 & 165.60 &  & 148.50 & 153.70 & 164.70 & 179.50 & 185.60 & \\
Fe & 16225.64 & 105.20 & 88.40 & 90.20 & 90.60 & 121.50 & 126.20 & 128.20 & 148.20\\
Fe & 16292.85 &  &  &  &  &  &  &  & \\
Fe & 16324.46 & 159.70 & 138.70 & 147.40 & 148.30 & 163.50 & 178.50 & 183.10 & 206.80\\
Fe & 16331.53 &  &  &  &  &  &  &  & \\
Ni & 16310.51 & 103.20 & 95.34 & 86.50 & 99.10 & 111.50 & 124.50 & 123.70 & 134.40\\
Ni & 16363.11 &  &  &  &  &  &  &  & \\
Ni & 16388.75 &  &  &  &  &  &  &  & \\
Ni & 16584.44 & 53.20 & 54.60 & 43.00 & 50.41 & 63.40 & 74.90 & 72.50 & 85.10\\
Ni & 16673.71 & 53.45 & 53.20 & 41.36 & 49.00 & 64.50 & 70.29 & 68.95 & 79.20\\
Ni & 16815.47 & 48.41 & 46.70 & 38.50 &  & 57.50 & 68.40 & 65.00 & 75.90\\
Mg & 15879.56 & 199.20 & 174.10 & 132.80 & 139.50 & 207.70 & 189.70 & 184.50 & 202.60\\
Mg & 15886.19 & 144.70 & 128.10 & 93.18 & 94.68 & 157.10 & 149.00 & 135.90 & 154.60\\
Al & 16718.98 &  & 318.80 & 293.90 & 307.80 & 392.70 & 358.50 & 339.70 & 354.60\\
Al & 16750.57 &  &  &  &  &  &  &  & \\
Al & 16763.35 & 235.70 & 202.90 & 184.60 & 187.20 & 248.60 & 247.50 & 245.20 & 281.50\\
Si & 15557.79 &  & 184.50 & 175.00 & 180.60 & 214.40 & 210.50 &  & 208.70\\
Si & 16060.02 & 255.20 & 243.20 & 230.70 & 231.30 & 255.50 & 275.30 & 253.20 & 275.00\\
Si & 16163.71 & 171.70 & 184.70 & 175.50 & 183.40 & 195.45 & 205.60 & 188.45 & \\
Si & 16215.71 & 226.70 & 227.60 & 214.50 & 227.60 & 249.50 & 258.50 & 237.80 & 260.20\\
Si & 16241.87 & 193.60 & 192.60 & 183.60 & 195.00 & 210.00 & 219.90 & 206.30 & 222.50\\
Si & 16828.18 & 145.70 & 159.50 & 151.60 &  & 163.40 & 184.30 & 166.30 & \\
Ca & 16150.76 & 128.50 & 101.50 & 70.66 & 81.30 & 115.30 & 121.60 & 112.70 & 134.90\\
Ca & 16155.27 & 64.10 & 48.65 & 30.10 &  & 58.50 & 69.60 & 57.60 & 72.08\\
Ca & 16197.04 & 192.50 &  & 115.30 &  & 171.45 & 169.50 &  & 185.50\\
Ti & 15602.84 & 35.55 &  &  & 16.50 & 31.20 & 34.55 & 43.60 & 50.39\\
Ti & 15715.57 &  &  &  &  &  &  &  & \\
Ti & 15543.78 &  &  &  & 72.74 & 99.50 &  &  & 155.20\\
\hline
\end{tabular}
\end{table*}

\begin{table*}
\centering
\caption{Equivalent width measurements for individual lines in m\AA\ for KIC9761625, KIC3455760, KIC9002884, KIC2142095, KIC8611114, KIC7594865, KIC1432587, KIC3658136.}
\label{EWs5}
\begin{tabular}{lccccccccc}  
\hline
\multicolumn{1}{c}{Element} &\multicolumn{1}{c}{$\lambda$ [\AA\ ]} &\multicolumn{1}{c}{KIC9761625} & \multicolumn{1}{c}{KIC3455760} & \multicolumn{1}{c}{KIC9002884}& \multicolumn{1}{c}{KIC2142095} &  \multicolumn{1}{c}{KIC8611114}& \multicolumn{1}{c}{KIC7594865}& \multicolumn{1}{c}{KIC1432587}& \multicolumn{1}{c}{KIC3658136}\\
\hline
Fe & 15176.72 &  &  &  &  &  &  &  & \\
Fe & 15301.56 &  &  &  &  & 63.50 &  &  & 86.35\\
Fe & 15394.67 &  &  &  &  &  &  &  & \\
Fe & 15395.72 &  &  &  &  &  &  &  & \\
Fe & 15485.45 &  &  &  &  &  &  &  & \\
Fe & 15524.30 &  &  &  &  &  &  &  & \\
Fe & 15534.26 &  & 170.25 &  & 122.70 & 150.30 &  & 169.30 & \\
Fe & 15648.52 &  &  &  &  &  &  &  & \\
Fe & 15652.87 &  &  &  &  &  &  &  & \\
Fe & 15733.51 &  &  &  &  &  &  &  & \\
Fe & 15761.31 &  &  &  &  &  &  &  & \\
Fe & 15774.07 & 159.50 &  & 165.60 & 135.80 & 160.10 & 145.20 & 171.40 & 178.30\\
Fe & 15609.04 &  &  &  &  &  &  &  & \\
Fe & 15904.35 &  &  &  &  &  &  &  & \\
Fe & 15906.04 &  & 206.40 & 202.30 & 156.40 & 180.30 & 172.20 & 207.10 & 209.20\\
Fe & 15934.02 & 87.28 & 89.79 & 87.43 & 52.18 & 71.60 & 71.10 & 88.20 & 95.20\\
Fe & 15952.63 & 48.29 & 47.55 & 46.30 & 22.80 & 37.60 & 36.00 & 48.20 & 55.20\\
Fe & 15964.87 &  &  &  &  &  &  &  & \\
Fe & 15980.73 &  &  &  &  &  &  &  & \\
Fe & 15997.74 &  &  &  &  &  &  &  & \\
Fe & 16040.65 &  &  &  &  &  &  &  & \\
Fe & 16102.41 & 209.90 & 240.70 & 226.10 & 185.30 & 203.80 & 196.50 & 226.10 & 233.40\\
Fe & 16180.93 &  & 168.50 & 159.40 & 119.30 & 147.50 & 141.45 & 168.40 & 173.40\\
Fe & 16198.51 & 192.20 &  & 210.50 & 153.20 & 179.80 & 171.80 & 214.80 & 214.80\\
Fe & 16225.64 &  & 143.40 & 134.90 & 94.95 & 121.30 & 116.00 &  & \\
Fe & 16292.85 &  &  &  &  &  &  &  & \\
Fe & 16324.46 & 190.50 & 207.60 & 210.20 & 145.80 & 175.30 & 167.80 & 213.30 & 211.50\\
Fe & 16331.53 &  &  &  &  &  &  &  & \\
Ni & 16310.51 & 135.40 & 136.10 & 139.30 & 90.07 & 109.80 & 106.20 & 134.10 & 145.20\\
Ni & 16363.11 &  &  &  &  &  &  &  & \\
Ni & 16388.75 &  &  &  &  &  &  &  & \\
Ni & 16584.44 & 85.05 & 80.59 & 84.80 & 46.90 & 58.80 & 57.45 & 81.50 & 93.20\\
Ni & 16673.71 & 82.55 & 78.27 & 75.50 & 46.50 & 58.70 & 57.75 & 74.50 & 82.50\\
Ni & 16815.47 & 80.30 & 72.50 & 76.85 & 41.22 & 51.46 & 52.28 & 73.30 & 80.50\\
Mg & 15879.56 & 198.20 & 210.10 & 198.90 & 151.30 & 149.70 & 155.60 & 182.50 & 176.10\\
Mg & 15886.19 & 153.50 & 161.50 & 151.80 & 102.90 & 105.20 & 110.10 & 137.70 & 132.50\\
Al & 16718.98 & 381.10 & 368.10 & 387.80 & 312.90 & 310.10 & 344.40 & 399.60 & 372.70\\
Al & 16750.57 &  &  &  &  &  &  &  & \\
Al & 16763.35 & 289.60 & 275.50 & 302.80 & 182.50 & 194.70 & 224.70 & 270.10 & 287.00\\
Si & 15557.79 &  &  &  & 175.70 &  & 196.90 &  & \\
Si & 16060.02 & 259.30 & 265.90 & 251.80 & 253.00 & 255.50 & 253.30 & 266.20 & 263.50\\
Si & 16163.71 & 204.70 & 194.70 & 196.30 & 165.50 & 192.80 & 193.10 & 211.40 & \\
Si & 16215.71 & 248.80 & 250.50 & 243.80 & 220.70 & 237.30 & 241.60 & 257.70 & 252.60\\
Si & 16241.87 & 220.10 & 211.90 &  & 180.60 & 204.30 & 213.10 & 225.50 & 224.10\\
Si & 16828.18 &  & 166.00 & 176.60 & 135.60 & 164.60 & 166.20 & 188.40 & 183.40\\
Ca & 16150.76 & 140.70 & 138.40 & 164.20 & 88.66 & 99.55 & 95.50 & 138.40 & 142.00\\
Ca & 16155.27 & 78.70 & 74.49 & 95.47 & 38.70 & 48.96 & 45.45 & 72.33 & 79.38\\
Ca & 16197.04 &  & 197.50 & 218.70 & 145.40 &  &  & 190.40 & 196.50\\
Ti & 15602.84 & 69.50 &  & 100.50 & 19.37 & 27.40 & 30.50 & 85.20 & 85.90\\
Ti & 15715.57 &  &  &  &  &  &  &  & \\
Ti & 15543.78 & 173.50 & 145.50 & 230.10 & 79.81 & 104.43 & 110.27 & 210.50 & 203.40\\
\hline
\end{tabular}
\end{table*}

\begin{table*}
\centering
\caption{Equivalent width measurements for individual lines in m\AA\ for KIC4446181, KIC4751953, KIC4661299, KIC4548530.}
\label{EWs6}
\begin{tabular}{lccccc}  
\hline
\multicolumn{1}{c}{Element} &\multicolumn{1}{c}{$\lambda$ [\AA\ ]} &\multicolumn{1}{c}{KIC4446181} & \multicolumn{1}{c}{KIC4751953} & \multicolumn{1}{c}{KIC4661299}& \multicolumn{1}{c}{KIC4548530}\\
\hline
Fe & 15176.72 &  &  &  & \\
Fe & 15301.56 & 44.94 & 52.20 & 34.45 & 58.69\\
Fe & 15394.67 &  &  &  & \\
Fe & 15395.72 &  &  &  & \\
Fe & 15485.45 &  &  &  & \\
Fe & 15524.30 &  &  &  & \\
Fe & 15534.26 &  &  &  & \\
Fe & 15648.52 &  &  &  & \\
Fe & 15652.87 &  &  &  & \\
Fe & 15733.51 &  &  &  & \\
Fe & 15761.31 &  &  &  & \\
Fe & 15774.07 &  &  &  & \\
Fe & 15609.04 &  &  &  & \\
Fe & 15904.35 & 125.70 & 128.50 &  & \\
Fe & 15906.04 & 148.80 & 148.10 &  & \\
Fe & 15934.02 & 53.20 & 54.86 & 38.80 & 62.05\\
Fe & 15952.63 &  & 28.02 & 19.55 & 34.47\\
Fe & 15964.87 &  &  &  & \\
Fe & 15980.73 &  &  &  & \\
Fe & 15997.74 &  &  &  & \\
Fe & 16040.65 & 142.10 & 145.50 & 127.80 & \\
Fe & 16102.41 & 164.30 & 162.10 & 145.30 & 188.60\\
Fe & 16180.93 & 118.60 &  & 94.03 & 129.40\\
Fe & 16198.51 & 146.10 & 149.10 & 121.50 & 160.80\\
Fe & 16225.64 & 90.40 & 92.79 &  & 110.10\\
Fe & 16292.85 & 80.81 & 82.10 &  & 92.35\\
Fe & 16324.46 & 142.90 & 145.40 & 121.40 & 159.80\\
Fe & 16331.53 &  &  &  & \\
Ni & 16310.51 & 102.10 & 98.90 & 91.40 & 98.85\\
Ni & 16363.11 & 159.90 & 152.50 & 143.60 & 151.80\\
Ni & 16388.75 &  &  &  & \\
Ni & 16584.44 &  &  &  & 51.91\\
Ni & 16673.71 & 55.40 &  & 49.58 & 51.60\\
Ni & 16815.47 & 51.45 & 51.22 & 40.89 & 49.35\\
Mg & 15879.56 & 169.70 & 158.70 & 134.00 & 177.60\\
Mg & 15886.19 & 122.70 & 113.50 & 97.94 & 131.30\\
Al & 16718.98 & 328.70 & 312.90 & 316.90 & 332.50\\
Al & 16750.57 &  &  &  & \\
Al & 16763.35 & 210.50 & 203.50 &  & 218.90\\
Si & 15557.79 & 186.50 & 187.40 & 188.70 & \\
Si & 16060.02 & 234.50 & 245.40 & 239.40 & 258.90\\
Si & 16163.71 & 184.75 & 185.50 & 178.80 & 196.70\\
Si & 16215.71 & 225.10 & 229.90 & 219.50 & 237.10\\
Si & 16241.87 & 198.30 & 196.50 & 200.70 & 207.40\\
Si & 16828.18 &  & 163.45 & 157.00 & 171.00\\
Ca & 16150.76 & 83.38 & 88.45 & 89.30 & 97.45\\
Ca & 16155.27 & 36.55 & 40.65 & 44.65 & 42.50\\
Ca & 16197.04 &  & 127.25 & 134.80 & 137.80\\
Ti & 15602.84 & 29.15 & 37.50 & 19.50 & 40.55\\
Ti & 15715.57 & 98.45 & 108.50 & 73.90 & 124.10\\
Ti & 15543.78 &  & 119.10 &  & 133.45\\
\hline
\end{tabular}
\end{table*}



\bsp	
\label{lastpage}
\end{document}